\documentclass[preprint,11pt]{JHEP3} % 10pt is ignored!

\JHEPspecialurl{http://jhep.sissa.it/JOURNAL/JHEP3.tar.gz}

\usepackage{epsfig,multicol,amsmath}

\usepackage{epstopdf}
\usepackage{rotating}
\usepackage{cite}

\def\beq{\begin{equation}}
\def\eeq{\end{equation}}
\def\bea{\begin{eqnarray}}
\def\eea{\end{eqnarray}}

\def\eq#1{{Eq.~(\ref{#1})}}
\def\fig#1{{Fig.~\ref{#1}}}

\newcommand{\bas}{\bar{\alpha}_S}
\newcommand{\as}{\alpha_S}

 \newcommand{\SP}{\langle \mid S^2 \mid\rangle}

\newcommand{\Lb}{\left(}
\newcommand{\Rb}{\right)}
\parskip=\itemsep               %?

\newcommand{\D}{\partial}
\newcommand{\h}{\frac{1}{2}}

\newcommand{\A}{{\cal A}}

\def\pom{{I\!\!P}}
\def\reg{{I\!\!R}}
\title{A QCD motivated model for soft interactions at high energies}
\author{\Large  E. Gotsman\thanks{Email:
gotsman@post.tau.ac.il.}\,, E. Levin\thanks{Email:
leving@post.tau.ac.il,
levin@mail.desy.de.}\,, U. Maor\thanks{Email:maor@post.tau.ac.il.}\, \,and\, J.S. Miller\thanks{Email:jeremymi@post.tau.ac.il.}\\
Department of Particle Physics, School of Physics and Astronomy\\
Raymond and Beverly Sackler Faculty of Exact Science\\  
Tel Aviv University, Tel Aviv, 69978, Israel}

\abstract{In this paper we  develop  an approach to soft scattering  processes at high
energies,
which is based on two mechanisms: Good-Walker mechanism for low mass diffraction
and multi-Pomeron interactions for high mass diffraction. The pricipal
idea,
 that allows
us to specify the theory for Pomeron interactions, is that the  so called
soft
 processes occur at rather short distances
 ($r^2 \propto 1 /<p_t>^2  \propto \alpha
'_\pom \approx 0.01 \,GeV^{-2}$), where  perturbative QCD is valid.  The
value
of the Pomeron slope $\alpha'_\pom $ was obtained from the fit to
 experimental data.
Using this theoretical approach we suggest a model that fits all soft data
in the ISR-Tevatron energy range, the
 total, elastic,
single and double diffractive cross sections,
 including $t$ dependence of the differential elastic cross section, and
the
  mass dependence of single
diffraction.
In this model we calculate the survival probability of  diffractive Higgs
production, and obtained a  value for this observable, which is smaller
than 1\% at the LHC energy range.}

\dedicated{PACS: 13.85.-t, 13.85.Hd, 11.55.-m, 11.55.Bq}
\preprint{TAUP -2878-08\\
{\tt 0805.2799 [hep-ph]}\\
\today}
\begin{document}
%%%%%%%%%%%%%%%%%%%%%%%%%%%%%%%%%%%%%%%%%%%%%%%%%%%%%%%%%
\section{Introduction}
%%%%%%%%%%%%%%%%%%%%%%%%%%%%%%%%%%%%%%%%%%%%%%%%%%%%%%%%%
\par
The goal of this paper is to construct a QCD motivated model for the 
strong interactions at high energy,
and  to elucidate
the model's predictions and implications at LHC and Cosmic Rays energies.

The difficulties and challenges of  a theoretical  approach to the strong
 interaction  are well known:  a qualitative   understanding of the
 confinement of quarks and gluons in QCD.  For high energy scattering the
 situation  is even more difficult than for the hadron structure, since 
 approximate 
methods such as QCD sum rules and/or effective theories, as well as the
lattice QCD approach, cannot be used to calculate  the high
energy amplitudes. Today, and for the past four decades,  the accepted
method of 
describing   
soft interactions at high energy, is  the phenomenology based on the
soft Pomeron and secondary Reggeons  (see Refs. \cite{COL,SOFT,LEREG} for
 details). All parameters related to the Pomeron and Reggeons, which are assumed to be
simple poles in the J-plane,  such as the
 intercepts and slopes of the trajectories, the vertices and their dependence
 on the impact parameter,   have to be deduced from the 
experimental data. However, the key problem of all approaches based on the soft
 Pomeron hypothesis, is that there is no theory which can specify what
kind of
 interactions between Pomerons
 have to be taken into account, as well as the values of the
multi-Pomeron vertices. Due to this problem we are doomed to build basically
 unreliable models, since the only criteria is that they describe
the experimental data (see
Refs. \cite{2CH,KMROLD,GLMLAST,KMRNEW}).

%%s a simple pole in the J-plane. Consequently, one has to also consider
%multi-Pomeron interactions for which we depend on models which are, essentially, parametrizations aimed at
%describing experimental data  (see Refs.[4-7,12] for details).

%    In a recent paper [6], we have attempted to overcome this difficulty by assuming [B +GLM]
%the soft Pomeron to be a partonic saturated limit of the DIS hard Pomeron, hence it is non-
%factorizable. This paper is more ambitious, in as much as we keep the taditional
%Regge-like features of the soft Pomeron using... }

In this paper we attempt to overcome this  difficulty by using
high
energy perturbative QCD approach.  At first sight, this appears 
unrealistic,  since the
 pQCD approach is based on the smallness of the running QCD coupling at 
short distances, while the high energy interaction is a  typical 
example of long distance non-perturbative QCD.  
 We wish to question this widely held prejudice. Indeed, the  only 
microscopic explanation 
for the Pomeron structure is given in the partonic approach \cite{GRIB}, 
in which
  the slope of the Pomeron trajectory is related to the mean transverse
 momentum of the exchanged partons
  ($\alpha'_\pom\,\propto 1/<p_t>$, where $< p_t> $ is the
mean parton momentum).
 The commonly held view in  high energy phenomenology, is
 that $\alpha'_\pom \,=\,0.25 \,GeV^{-2}$\cite{DL},   which implies that in 
the Pomeron the typical momenta of partons are  high (especially if 
you compare this value with
the value of  $\alpha'_{\reg}\,= \,1\, GeV^{-2}$ for the secondary Reggeons).
 In the next section, we extend our output up to the Planck mass,
 and  will show that the
fit to the experimental data based on our model with power-like
 dependence of the Pomeron-proton vertices on energy, validates the consistency 
of our calculations with
unitarity. The 
 momentum transferred $t$  behaviour leads to  $\alpha'_\pom  <  0.02 $
 $GeV^{-2}$ . This result, together with the fit of Ref. \cite{KMRNEW},
where
 the data were fitted with  $\alpha'_\pom  = 0$, lends support to
 our  assumption that, the typical parton momentum is  large
 (approximately $<p_t> = 1/\sqrt{\alpha'_\pom} \geq 7 \,GeV$). Therefore,
 the running QCD coupling $ \as =  \pi/b \ln\Lb <p^2_t>/\Lambda^2_{QCD}
 \Rb \,\ll\,1$ (approximately  0.18), and we can consider it as
 our small parameter, when applying perturbative  QCD  estimates to
 the Pomeron-Pomeron interaction vertices.

The theoretical scheme  that emerges from such an approach will be
 discussed in section 3. Using  perturbative  QCD
we select the essential Pomeron-Pomeron vertices, and develop the method 
for summation of all these diagrams. We also give the partonic
interpretation
 of our approach, and develop an approximate method for the solution of
the
 long standing problem of the high energy asymptotic behaviour of the 
scattering amplitude. Based on  this analysis  we derive formulae for
all cross sections taking into account the Pomeron interactions.

In section 4 we fit the available experimental data using our formalism.
This fit  is based on an updated data base
which includes the published $p$-$p$ and $\bar{p}$-$p$ data points of
$\sigma_{tot}$, the integrated values of $\sigma_{el}$, $\sigma_{sd}$,
$\sigma_{dd}$ and the forward elastic slope $B_{el}$, in the ISR-Tevatron
energy range. $\rho=\frac{Re\,a_{el}(t=0,s)}{Im\,a_{el}(t=0,s)}$ and
the forward slope of the SD and DD final states are  predictions of the
model. 
  The additional 
 data that we have used to validate our present model,
are the $t$-dependence of the  differential elastic cross section, and the mass
dependence
of the single diffractive production. We successfully describe this data.
Based on this fit  we give  reliable
predictions of the quantities to be measured at  the LHC c.m.
energy of $ 14 \, TeV$.
Our output also covers  the broad Cosmic Rays energy range up to the GZK
limit.

The fifth section is devoted entirely to a  calculation of the survival
probability ($\SP$) for  exclusive central diffractive Higgs production
 at the LHC, and a discussion on
the reliability of these
calculations. We confirm the tendency in which $\SP$ becomes
 small (less that 1\%),   noticed in our previous
paper \cite{GLMLAST}. From a practical point of view, this is the most
salient feature of this paper.

In the conclusions, we compare critically our results with other approaches,
 and summarize our findings.
 We list a few  experimental signatures, which should enable us to
differentiate between alternative theoretical options and phenomenological models.
{\boldmath
\section{The two channel model and  the value of $\alpha'_\pom$}}
\subsection{GLM two channel model}
%%%%%%%%%%%%%%%%%%%%%%%%%%%%%%%%%%%%%%%%%%%%%%%%
\par
The GLM two channel model has been described in our previous publications
(see Refs.\cite{2CH,SP2CH,heralhc,GLMLAST,SP3P}, and references therein).
In this formalism, diffractively produced hadrons at a given vertex are
considered as a single hadronic state
described by the wave function $\Psi_D$, which is orthonormal
to the wave function $\Psi_h$ of the incoming hadron (proton in the case of
interest), $<\Psi_h|\Psi_D>=0 $.
We introduce two wave functions $\psi_1$ and $\psi_2$ that diagonalize the
2x2 interaction matrix ${\bf T}$
\beq \label{2CHM}
A_{i,k}=<\psi_i\,\psi_k|\mathbf{T}|\psi_{i'}\,\psi_{k'}>=
A_{i,k}\,\delta_{i,i'}\,\delta_{k,k'}.
\eeq
In this representation the observed states are written in the form
\beq \label{2CHM31}
\psi_h=\alpha\,\psi_1+\beta\,\psi_2\,,
\eeq
\beq \label{2CHM32}
\psi_D=-\beta\,\psi_1+\alpha \,\psi_2\,,
\eeq
where, $\alpha^2+\beta^2=1$.
Using \eq{2CHM}, we can rewrite the unitarity constraints in the form
\beq \label{UNIT}
Im\,A_{i,k}\left(s,b\right)=|A_{i,k}\left(s,b\right)|^2
+G^{in}_{i,k}(s,b),
\eeq
where $G^{in}_{i,k}$ is the contribution of all non diffractive inelastic
processes,
i.e. it is the summed probability for these final states to be
produced in the scattering of particle $i$ off particle $k$.

A simple solution to \eq{UNIT} has the same structure as in a single channel
formalism, 
\beq \label{2CHM1}
A_{i,k}(s,b)=i \Lb 1 -\exp\Lb - \frac{\Omega_{i,k}(s,b)}{2}\Rb\Rb,
\eeq
\beq \label{2CHM2}
G^{in}_{i,k}(s,b)=1-\exp\Lb - \Omega_{i,k}(s,b)\Rb.
\eeq
From \eq{2CHM2} we deduce, that the probability that the initial
projectiles
$(i,k)$ reach the final state interaction unchanged, regardless of the initial
state rescatterings, is
$P^S_{i,k}=\exp \Lb - \Omega_{i,k}(s,b) \Rb$.
\par
For the opacities $\Omega_{i,k}$ we use   the  expression
\beq \label{OMEGA}
\Omega_{i,k}\Lb s,b \Rb\,=\,g_{i} \,g_{k}
\Lb \frac{s}{s_0} \Rb^{\Delta_\pom}\,S\Lb b ;m_i,m_k;  \alpha'_\pom \ln (s/s_0) \Rb
\eeq
which   differs from the expression that we used in our previous
models. The profile function\\
$S\Lb b , \alpha'_\pom ; m_i,m_k; \ln (s/s_0) \Rb$ at $s =s_0$,
corresponds
 to the power-like behaviour of the Pomeron-hadron vertices,
\bea
&&S\Lb b ; m_i, m_k; , \alpha'_\pom\ln (s/s_0) =0 \Rb \,\,=\,\,\int\,\frac{d^2 q}{(2 \pi)^2} \,g_i(q)\,g_k(q)\,e^{
 i \vec{q}_\perp \cdot \vec{b}} \label{SB} ,\\
&& \mbox{with a normalization} \,\int\,d^2 b\,S\Lb b ; m_i, m_k; , \alpha'_\pom\ln (s/s_0) =0 \Rb\,\,=1\,.
\notag
\eea
\par
In this paper we choose 
\beq  \label{GQ}
g_i(q)\,\,=\,\,\frac{1}{ (1\,\,+\,\,q^2/m^2_i)^2}
\eeq
The  arguments we  list  in  favour of this choice are: (i)  pQCD leads to
$ g_i(q)\,\,\rightarrow\,\,\as^2(q)/q^4$ \cite{BRLE}; (ii)
 in some models (for example in the constituent quark model)
 $ g_i(q) $ is equal to the electro-magnetic form factor of the proton, which 
has the form of \eq{GQ};
and (iii) \eq{GQ}  reproduces the experimental  elastic differential  cross section for p-p (p-${\bar p})$
data  in the range of $t$, up to 1  to 1.5 $\,GeV^2$ \cite{DL}, while a
Gaussian parametrization
   fits the data only for very small $t \leq 0.1\,GeV^2$. 

Using \eq{SB} we obtain for $ S\Lb b ; m_i, m_k; , \alpha'_\pom\ln (s/s_0) =0 \Rb$,
\bea
&&\frac{1}{ (1\,\,+\,\,q^2/m^2_i)^2 }\times \frac{1}{ (1\,\,+\,\,q^2/m^2_k)^2}\,\,\,\Longrightarrow\,\,\,S\Lb b ; m_i, m_k; , \alpha'_\pom\ln (s/s_0) =0 \Rb\,\,=\,\, \,\label{SDB}\\
&& =\,\,\frac{m^3_i\,m^3_k}{4 \pi\,(m^2_i - m^2_k)^3} \,\,\left\{
4 m_i\,m_k \Lb K_0\Lb m_i b \Rb \,-\,K_0\Lb m_k b \Rb \Rb \,\,+\,\,(m^2_i - m
^2_k) b\,\Lb m_k K_1\Lb m_i b \Rb \,+\, m_i K_1\Lb m_k b \Rb \Rb \right\}.
 \notag
\eea 
For energies $s > s_0$, we need to take into account the observed
shrinkage of the diffraction peak. To this
end we
 replace  $g_i(q)\,g_k(q)$ by $ g_i(q)\,g_k(q)\,\exp\Lb - \alpha'_\pom\ln (s/s_0) q^2 \Rb$ in 
\eq{SB}.
  To simplify our calculations we replace
\beq \label{MS}
m^2_i \,\,\,\Longrightarrow\,\,\,m^2_i(s)\,\,\equiv\,\,\,\frac{m^2_i}{1\,\,+\,\, \alpha'_\pom\ln
(s/s_0) /4m^2_i}.
\eeq
 It is easy to check  that   $ g_i(q; m_i)\,g_k(q; m_k)\,\exp\Lb -
\alpha'_\pom\ln (s/s_0) q^2 \Rb$ and 
$ g_i(q; m_i(s))\,g_k(q; m_k(s))$ have the same behaviour for $  \alpha'_\pom\ln (s/s_0) q^2\,\ll\,1$.  When 
$  \alpha'_\pom\ln (s/s_0) q^2\,\gg\,1$ these two expressions are different. Note that
the  Regge factor $\exp\Lb - \alpha'_\pom\ln (s/s_0) q^2 \Rb$ cannot be justified in this kinematic region.
 In the region of  $  \alpha'_\pom\ln (s/s_0) q^2\,\sim\,1$  it is
preferable to use the Regge factor.
 We assume that $ \alpha'_\pom$ is  small enough, so
 that this region  gives a negligible
contribution to all  experimental observables.

In general,  we have to consider four
possible re-scattering processes in \eq{2CHM}. However, in the case of
$p$-$p$ ( $\bar p$-$p$) the two non-diagonal amplitudes are equal
$A_{1,2}=A_{2,1}$, and we end  up with  three rescattering amplitudes.
These amplitudes are presented
in our two channel formalism in
the following form\cite{2CH,SP3P,heralhc}
\beq \label{EL}
a_{el}(s,b)=
i\{\alpha^4A_{1,1}+2\alpha^2\beta^2A_{1,2}+\beta^4\A_{2,2}\},
\eeq
\beq \label{SD}
a_{sd}(s,b)=
i\alpha\beta\{-\alpha^2A_{1,1}+(\alpha^2-\beta^2)A_{1,2}+\beta^2A_{2,2}\},
\eeq
\beq \label{DD}
a_{dd}=
i\alpha^2\beta^2\{A_{1,1}-2A_{1,2}+A_{2,2}\}.
\eeq
It should be stressed that in this approach diffraction dissociation,
appears as an outcome of the  Good and Walker mechanism\cite{GW} (G-W) or,
in other words, 
the elastic, single diffraction and double diffraction processes occur due to  
elastic scatterings
of $\psi_{1}$ and $\psi_{2}$, the
correct degrees of freedom.
\par
The corresponding cross sections are given by
\beq \label{XST}
\sigma_{tot}(s)=2\int d^2 b \,a_{el}\Lb s,b\Rb,
\eeq
\beq \label{XSEL}
\sigma_{el}(s)=\int d^2 b \,|a_{el}\Lb s,b\Rb|^2,
\eeq
\beq \label{XSSD}
\sigma_{sd}(s)=\int d^2 b \,|a_{sd}\Lb s,b\Rb|^2,
\eeq
\beq \label{XSDD}
\sigma_{dd}(s)=\int d^2 b \,|a_{dd}\Lb s,b\Rb|^2.
\eeq
\par

{\boldmath
\subsection{$\alpha'_\pom \,\,\longrightarrow\,\,0$}
}

Using \eq{2CHM} - \eq{OMEGA} we  fit the experimental data, so as
to find the
 value of 
$\alpha'_\pom$. However,  most of the data are available at rather low energies $W= \sqrt{s}
= 20$ to $ 600 \,GeV$, where the contributions of the secondary Reggeon
have to be included.
For this we replace $ \Omega_{i,k}\Lb s,b \Rb$ given by \eq{OMEGA}  by the
sum of Pomeron and Reggeon contributions,
\beq \label{OMPR}
\Omega_{i,k}\Lb s,b \Rb\,\,\,\,=\,\,\,\Omega^\pom_{i,k}\Lb s,b \Rb\,\mbox{( see \eq{OMEGA})}\,\,\,+\,\,\,\Omega^{\reg}_{i,k}\Lb s,b \Rb,
\eeq
with
\beq \label{OMR}
\Omega^{\reg}_{i,k}\Lb s,b \Rb\,\,=\,\,\frac{g^{\reg}_{i}\,g^{\reg}_k\,\eta\,\Lb \frac{s}{s_0}\,\Rb^{\Delta_{\reg}}}{\pi R^2_{i,k}\Lb s \Rb}
\exp \Lb - \frac{b^2}{ R^2_{i,k}\Lb s \Rb}\Rb.
\eeq
 The  signature factor is
\beq \label{SGN}
\eta\,\,\,=\,\,\,\frac{1 \pm e^{ i \pi \alpha_{\reg}(q)}}{\sin\Lb\pi \alpha_{
\reg}(q)\Rb}
\eeq
corresponding to a Reggeon trajectory
$\alpha_{\reg}(q)\,=\,1 + \,\Delta_{\reg} \,+\,\alpha'_{\reg}\,q^2$ ,
and
\beq \label{radius}
R^2_{i,k}\Lb s \Rb = R^2_{0,i}+ R^2_{0,k}+4\alpha'_{\reg}\ln(s/s_0),
= R^2_{0;i,k} + 4 \alpha^{'}_{\reg} \ln(\frac{s}{s_{0}}).
\eeq
$g^R_i$,$  R^2_{0,i}$ , $\Delta_{\reg}$ and $\alpha^{'}_{\reg}$ are fitted parameters.
 For the Regge sector we know the natural values for $\Delta_{\reg}$ and
 $\alpha^{'}_{\reg}$ from the behaviour of the Regge trajectory at
 $q^2 <0$ ($t > 0$) ,  since all hadrons lie on these trajectories,
  $ \Delta_{\reg} \approx - 0.5$ and $ \alpha'_{\reg} \approx
1\,GeV^{-2}$.

 \eq{OMPR} - \eq{radius} specify our model.  In this
 model we neglect the interactions between Pomerons, but include
eikonal
type rescatterings. This  means that we assume only  a  G-W
origin for all
 quasi-elastic processes.  The fitted parameters are shown in Table 1.
The quality of the fit is  illustrated in \fig{xst} and \fig{bel}, where
 our fit is shown by dashed lines.   The fit is very good with
$\chi^2/d.o.f. = 0.87$.
However, this parametrization fails to fully  describe the cross sections
for single and
 double diffractive production, yielding values for these cross sections
which are
approximately two times smaller than the experimental values.

From this fit we have two  conclusions:\, i) we cannot describe the
diffractive production in the framework of the  G-W  mechanism if we
assume
an
exponential parameterization for the proton profile function; and (ii)
the value of
 $\alpha^{\prime}_{\pom}$  turns out to be very small $\alpha^{\prime}_{\pom} =
 0.012 \,GeV^{-2}$. As we have discussed in the introduction, we conclude
from the
smallness of $\alpha^{\prime}_{\pom} $,  that the hard  processes which
occur at short distances  are responsible for the Pomeron structure.

\TABLE[ht]{
\begin{tabular}{|l|l|l|l|l|l|l|}
\hline
 $ \Delta_\pom $ & $\beta$ &  $\alpha^{\prime}_{\pom}$& $g_1$ &  $g_2$ & $m_1$ &
$m_2$  \\ \hline
0.120 & 0.46 & 0.012 $GeV^{-2}$ & 1.27 $GeV^{-1}$ & 3.33 $GeV^{-1}$ & 0.913 $GeV$& 0.98 $GeV$
\\ \hline
$ \Delta_\reg$ & $\beta$ &  $\alpha^{\prime}_{\reg}$& $g^{\reg}_1$ &  $g^{\reg}_2$ & $ R^2_{0,1}$& $\chi^2/d.o.f.$
 \\ \hline
-0.438 & 0.46 & 0.60 $GeV^{-2}$ & 4.0 $GeV^{-1}$ & 118.4 $GeV^{-1}$ & 4.0 $GeV^{-2}$ & 0.87
\\ \hline
\end{tabular}
\caption{Fitted parameters for two channel (eikonal) model}
\label{t1}}

%%%%%%%%%%%%%%%%%%%%%%%%%%%%%%%%%%%%%%%%%%%%%%%%%%%%%%%%%%%%%%%%%%%%%%%%
\DOUBLEFIGURE[ht]{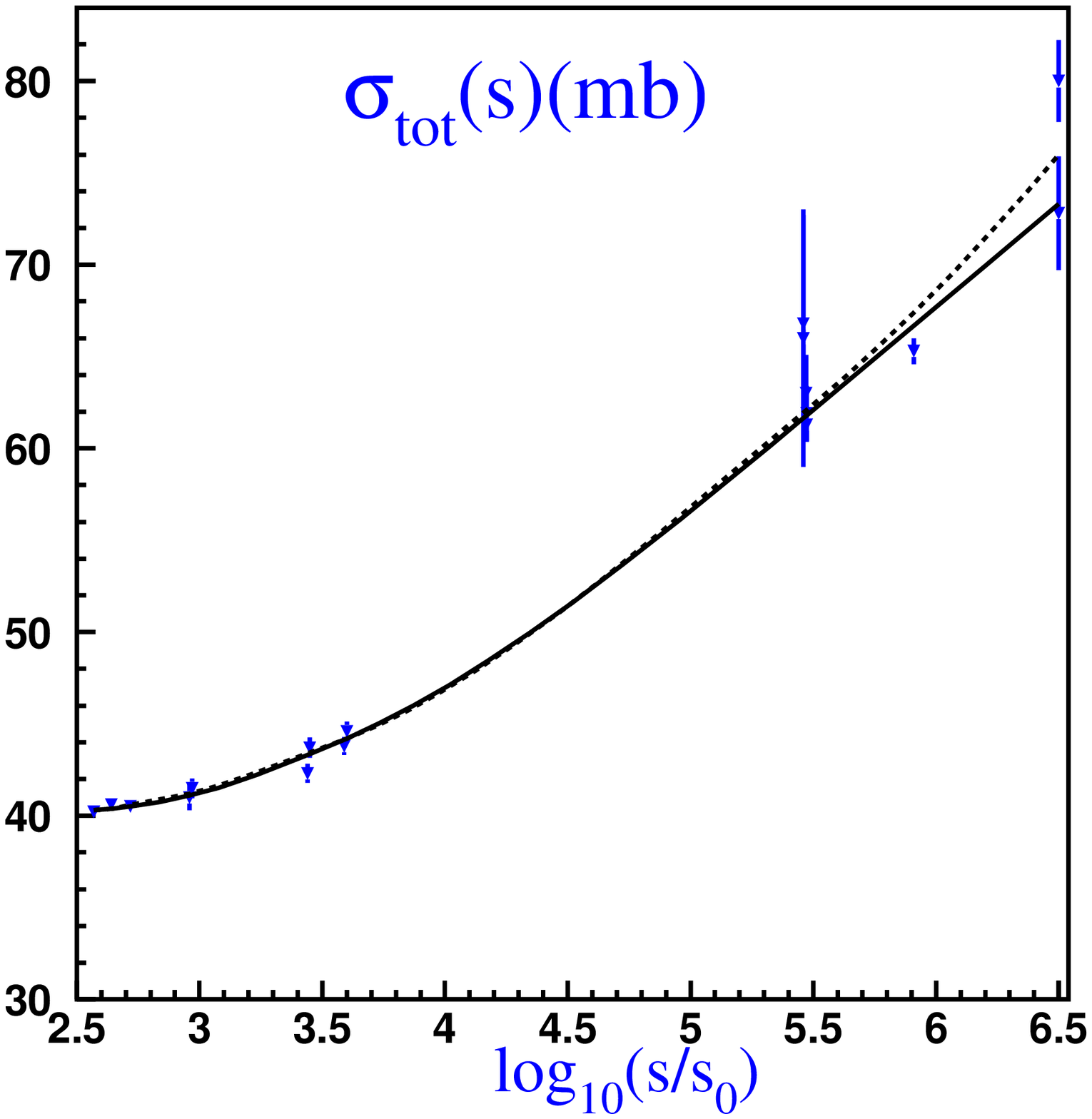,width=85mm,height=75mm}{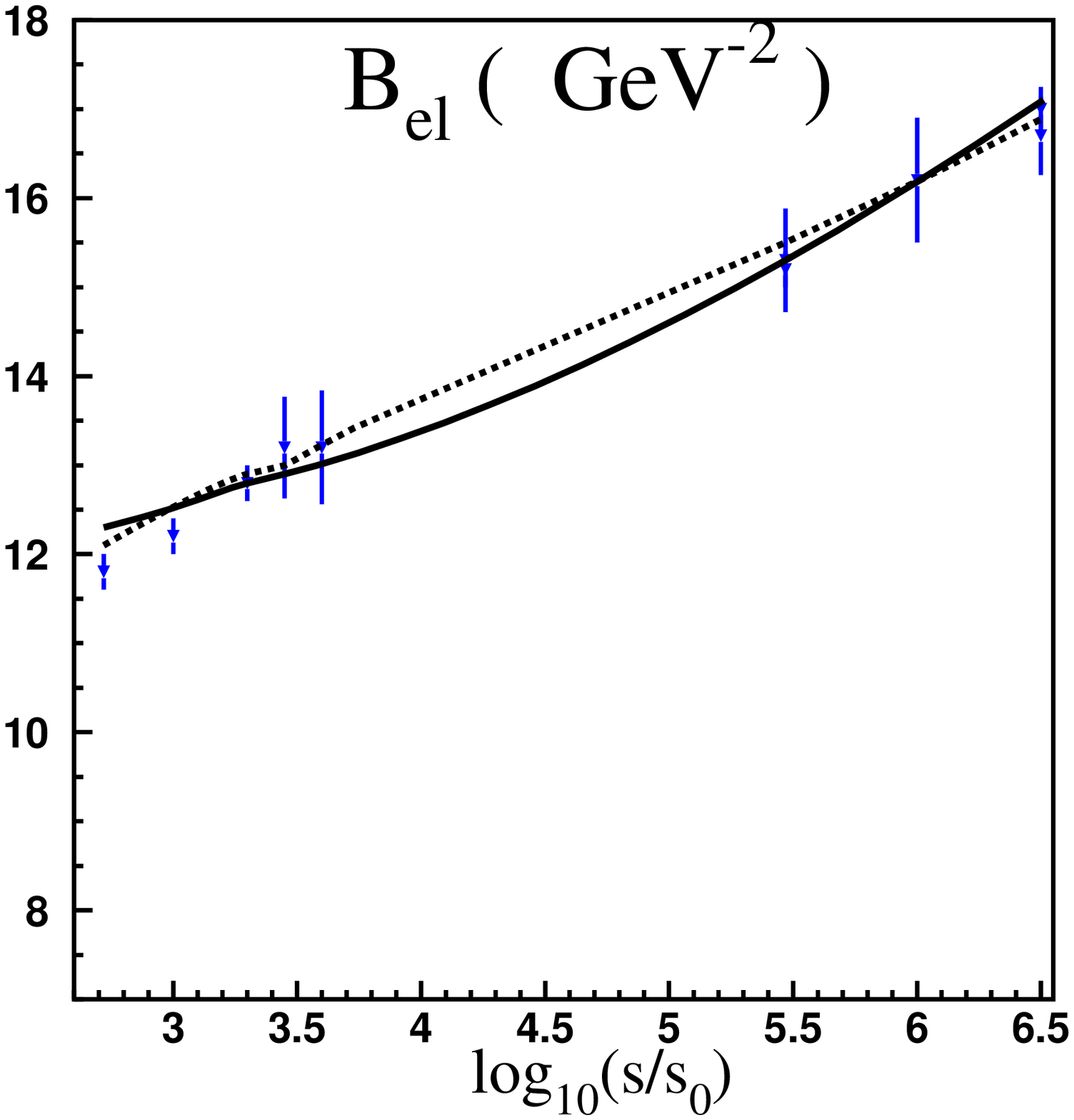,width=85mm,height=75mm}
{Energy dependence of $\sigma_{tot}$. The solid line shows the fit with taking into account all Pomeron interactions while the dashed line corresponds to two channel (eikonal) model.
\label{xst}}
{Energy dependence of the slope for the differential elastic cross section. All notation are the same as in \protect\fig{xst}
\label{bel}}
%%%%%%%%%%%%
\section{ Pomerons Interactions}
\subsection{QCD input}
Our main hypothesis is  that the typical distances for a Pomeron exchange
are
short, and we can use pQCD as a guide  for building a theory for Pomeron
interactions,
  based on the small value of $\alpha^{\prime}_{\pom}$,  the
Pomeron slope
 that we obtained in the previous section.  We  recall
that the value of  $ \alpha^{\prime}_{\pom}$ in the parton model \cite{GRIB} can be written as
\beq \label{ALHPR}
R^2(s)\,\,=\,\,\langle b^2 \rangle_n \,\,\,=\,\,\,\langle  \Delta b^2 \rangle\,n
\,\,\,= \,\,\frac{4}{<p^2_t>}\,\rho\,\ln(s/s_0)\,\,=\,\, \alpha^{\prime}_{\pom}\ln(s/s_0)
\eeq
where
$n$ is the number of partons at rapidity  $ Y  = \ln(s/s_0)$
and
 $\rho$ is theb density of partons in a unit of
rapidity.  In \eq{ALHPR} the radius of interaction, is determined as the average distance in impact parameter space that  a `wee' parton can reach at given enegy.

In \eq{ALHPR} $ R^2(s)$ is the radius of interaction at energy $W=
\sqrt{s}$,
 which is the distance in  impact parameter $b$ space, that is reached
by
 partons in a two dimensional diffusion,  after $n$ steps.
$\langle \Delta b^2 \rangle$ is the mean displacement during  one diffusion
 step, which can be estimated using the uncertainty principle as
 $\langle \Delta b^2 \rangle \,=\,4/<p^2_t>$. $<p_t>$  is the average parton
  transverse momentum which we wish to evaluate. The number of diffusion steps
 is equal to the number of partons, since at each step one  parton is
emitted. At a given energy the average number of emitted
partons is equal to $\rho \ln(s/s_0)$, where $\rho$ is the parton density
in units of rapidity.
Comparing $n$ with the hadron multiplicities, we  conclude that $\rho
\approx
  1 $ or larger. Therefore,
\eq{ALHPR}  leads to $<p^2_t \,\geq \, 20 \,GeV$ for $ \alpha^{\prime}_{\pom}  = 0.01\,GeV^{-2}$ . The typical value of the running QCD coupling, which corresponds to this value of the parton momentum, is
$\as \approx 0.15$, which is small enough to be used in pQCD estimates.
\par
 In our
procedure we only  use  general results of the perturbative QCD
approach
 to high energy scattering \cite{QCDHE}.
Consequently:

1.) In the leading order approximation of pQCD,  only a
Pomeron splitting into two Pomerons and two Pomeron merging into one
Pomeron,
 should be taken into account \cite{BART,BRN}, while all other vertices
are
 small.  Therefore, using this  input from pQCD, we
restrict ourselves by summing Pomeron diagrams with  triple Pomeron
vertices only.

2)   Since $4 \alpha^{\prime}_{\pom}ln(s)\,\, \ll\; 1$  over the
entire kninematical range, we have investigated, we
 can neglect the value of $
\alpha^{\prime}_{\pom}
$, and consider the theory with $ \alpha^{\prime}_{\pom}\,\,=\,\,0$.

3)  We  obtain the anticipated values for all ingredients of
the   Pomeron
interaction approach: the intercept of the Pomeron $\Delta_\pom$ above unity
$\,\propto \,\as$,  and the triple Pomeron vertex
coupling $g_{3\pom}\,\propto\,\as^2$.

The theory that includes all the above ingredients  can
be formulated in functional integral form\cite{BRN},
\begin{equation} \label{FI}
Z[\Phi, \Phi^+]\,\,=\,\,\int \,\,D \Phi\,D\Phi^+\,e^S \,\,\,\mbox{with}\,S \,=\,S_0
\,+\,S_I\,+\,S_E\,,
\end{equation}
where $S_0$ describes free Pomerons, $S_I$ corresponds to their mutual interaction
and $S_E$ relates to the interaction with the external sources (target and
projectile).  Since  $ \alpha^{\prime}_{\pom}\,=\,0$, $S_0$ has the form
\begin{equation} \label{S0}
S_0\,=\,\int d Y \Phi^+(Y)\,\left\{ -\,
\frac{d }{d Y} \,\,
+\,\,\Delta\,\right\} \Phi(Y).
\end{equation}
$S_I$ includes only triple Pomeron interactions and has the form
\beq \label{SI} S_I\,=\,g_{3\pom} \int d Y\,\left\{
\Phi(Y)\,
\Phi^+(Y)\,\Phi^+(Y)\,\,+\,\,h.c. \right\}. \eeq 
For $S_E$
we have local interactions both in rapidity and in impact parameter space,
 \begin{equation} \label{SE}
S_E\,=\,-\,\int dY \sum_{i=1}^2\,\left\{
\Phi(Y)\,g_i(b)\,\,+\,\,\Phi^+(Y)\, g_i(b)
\right\},
\end{equation}
where $g_i(b)$  stands for the interaction vertex with the hadrons at
fixed $b$.

In the next sections, after specifying this theory, we solve it. 
Indeed, this theory as any theory of  Pomeron interactions, is written
 in  such a way that an  high energy amplitude satisfies  $t$-channel
unitarity. However,  $s$-channel unitarity remains a problem.
 In the next subsection we will change the interaction term ($S_I$), in 
 such a way, that the theory will have a clear partonic interpretation, 
and
 will, also, satisfy  $s$-channel unitarity.

\subsection{Generating function and a partonic interpretation}

To find a reformulation  given by the  functional of 
\eq{FI},  we consider a system of partons\footnote{The partons for high 
energy QCD are the colourless dipoles, as was shown in Ref. \cite{MUCD}.}
 that can decay and merge:  one parton to two partons, and two partons 
into 
one parton, with probabilities $\Gamma(1 \to 2)$ and  $ \Gamma(2 \to 1)$,
 respectively.
For such a system of partons, we can write a simple equation. Indeed,
let $P_n(y)$ be the probability to find 
$n$-parton (dipoles) with rapidity $y$ in the wave function of the fastest (parent)
 parton (dipole),  moving with rapidity $Y\,>\,y$.
For $P_n(y)$, we can easily write down a recurrence equation (see
Refs. \cite{GRPO,LELU})
\beq \label{PNEQ}
-\,\frac{\partial\,P_n(y)}{\partial \,y}\,\,=\,\,\Gamma(1 \to 2)\,
 \left\{ -\,n\,P_n\,+\, \,(n-1)\,P_{n-1}  \right\}\,\,\,+\,\,\,\Gamma(2 \to 1)
 \, \left\{ -\,n\,(n - 1)\,P_n\,+\, \,(n+1)\,n\,P_{n+1}  \right\}.
\eeq
In each bracket the first term on the r.h.s., 
can be viewed as a probability of a dipole annihilation 
in the  rapidity range $( y $ to $ y - dy )$ (death  term). The second is a
probability to  create one extra  dipole (birth term). Note the
negative sign 
in front of $\partial P_n(y)/\partial y$. It appears due to our choice of the
rapidity evolution which starts at the largest rapidity $y=Y$,
of the fastest dipole and then decreases.
The first two terms are responsible for the process of  parton decay, 
while the last two terms describe the
contribution of partons merging.

It is useful to introduce the generating function \cite{MUCD,LALE,LELU} 
\beq   \label{Z}
Z(y,\,u)\,\,=\,\,\sum_n\,\,P_n(y)\,\,u^n\,
\eeq
At  rapidity $y\,=\,Y$ 
there is only one fastest parton (dipole), which is $P_1(y\,=\,Y)\,=\,1$,
while $P_{n>1}(y\,=\,Y)\,=\,0$. 
This is the initial condition for the generating
function
\beq \label{INC1}
Z(y\,=\,Y)\,=\,u\,.
\eeq
At $u =1$ 
\beq \label{INC2}
Z(y,\,u\,=\,1)\,\,=\,\,1,
\eeq
which follows from the physical meaning of $P_n$ as a  probability.

\eq{PNEQ} can be rewritten as an  equation in partial derivatives
for the generating function $Z(y,u)$,
\beq \label{GFEQ}
\,\,-\frac{\partial\,Z(y,\,u)}{\partial\, y}\,\,
=\,\,-\,\Gamma(1 \to 2)\,u\,(1\,-\,u) 
\,\,\frac{\partial\,Z(y,\,u)}{\partial\, u}\,\,\,+\,\,\,\Gamma(2 \to 1)\,u\,(1\,-\,u) \,\,\frac{\partial^2\,Z(y
,\,u)}{\partial^2\, u}.
\eeq
The description of the parton system given by \eq{GFEQ}, is equivalent to
the path integral of \eq{FI}.
Indeed,  the general solution of \eq{GFEQ} has a form
\beq \label{TMH1}
Z(Y;u)\,\,=\,\,e^{H(u)\,(Y - Y_0)}Z(Y_0;u),
\eeq
with the operator $H$ defined as
\beq \label{TMH2}
H(u)\,\,=\,\,-\, \Gamma(1 \to 2)\,u (1 - u)
\,\frac{\partial }{\partial u}+\,\Gamma(2 \to
1)\,u (1 - u)\,\frac{\partial^2 }{(\partial u)^2},
\eeq
and
\beq \label{TMZ0}
Z(Y_0;u)\,\,=\,\,e^{g(b)( u -1 )}.
\eeq

We introduce operators of creation ($a^+$) and annihilation ($a$)
that satisfy
$[\hat{a} ,\hat{a}^+] = 1$,
at fixed  Y.

\beq \label{TMH3}
\hat{a}\, =\,\frac{\D}{\D u}\, , \, \hat{a}^+ \,=\,u \,.
\eeq
In this formalism
\beq \label{TMH4}
 H\,\,=\,\,-\,\Gamma(1 \to 2)\,\hat{a}^+\,( 1 - \hat{a}^+)\,\hat{a} \,\,+\,\,\Gamma(2 \to
1)\,\hat{a}^+\,( 1 - \hat{a}^+)\,
\hat{a}^2,
\eeq
and the initial state at $Y=Y_0$ is defined as
\beq \label{TMIS}
|Y_0>\,\,=\,\,e^{g(b)( \hat{a}^+ \,-\,1)} |0>,
\eeq
with the vacuum defined as $\hat{a} |0> = 0$.

The theory with the Hamiltonian of \eq{TMH4}, has  an  equivalent
description using the  path integral of
\eq{FI} (see the detailed derivation in Ref. \cite{KLP})  with

\begin{eqnarray}
  S\,&=&\,\int \left( \Phi^+  \frac{d}{dY} \Phi + H ( \Phi^+ +1, -\Phi ) \, \right)
dY \,\, =\, \,\int d Y\,\,\left( \Phi^+  \frac{d}{dY} \Phi \,\,\right. \label{TMHAC} \\
 &-& \left.
\Gamma(1\rightarrow2) \Phi^+\Phi
+\Gamma(1\rightarrow2) \Phi^+(\Phi)^2
+\Gamma(2\rightarrow 1)( \Phi^+)^2\Phi
-\Gamma(2\rightarrow 1) (\Phi^+)^2\,(\Phi)^2
 \, \right).   \nonumber
\end{eqnarray}
Comparing \eq{TMHAC} with \eq{S0} - \eq{SE}, one can see that they are
similar if we put $
\Gamma( 1 \to 2) = \Delta$ and $ \Gamma(2 \to 1) \,=\,g^2_{3\pom}/\Delta$. However, \eq{TMHAC} has an additional term ($-\Gamma(2\rightarrow 1) (\Phi^+)^2\,(\Phi)^2$ )  which describes the two Pomeron to two
Pomeron transition (four Pomeron interaction).   This term ensures
 that our approach satisfies  $s$-channel unitarity (see
Ref.\cite{BOMU}),
 where it is shown that without this term
there is no probabilistic interpretation of the Pomeron interaction
theory,
 and  $s$-channel unitarity is violated.  We also need to renormalize
 $\Phi^+\,\to \,(g_{3\pom}/\Delta)\,\Phi^+$ and
$\Phi \,\to \, (\Delta/g_{3\pom})\,\Phi$.

 Using the functional $Z$,
we  find the scattering amplitude \cite{K}, using the following formula:
\beq \label{AM}
N\Lb Y\Rb\,\,\,\equiv\,\,\,\mbox{Im} A_{el}\Lb Y\Rb \,\,\,=\,\,\,\sum^{\infty}_
{n =1}\frac{(-1)^n}{n!}\,\,\,\frac{\partial^n\,Z(y,\,u)}{\partial^n\, u}|_{u =1
}\,\gamma_n(Y=Y_0,b),
\eeq
where $\gamma_n(Y=Y_0,b)$ is the scattering amplitude of $n$-partons (dipoles)
at low energy. These amplitudes
 depend on the impact parameters which are the same for all $n$ partons,
since
$\alpha'_\pom\,\,=\,\,0$, and we
  neglect the diffusion of partons  in   impact parameter space.
 \eq{AM}  corresponds to the partonic approach \cite{GRIB,FEY},  in which
a high energy scattering  can be viewed as a two stage process. The first stage is a development
of the partonic wave function, which we consider  by introducing the
generating
function $Z$. In \eq{AM},    the second stage is
 the interaction of the lowest energy partons
(`wee' partons) with
the target, which is described by
the amplitudes
 $\gamma_n(Y=Y_0,b)$.  Assuming that there are no correlations between the
interacting partons (dipoles) at low energy, we can consider
$\gamma_n(Y=Y_0,b)\,\,=\,\,\gamma^n_1(Y=Y_0,b)$ \cite{K}, which we
 include, choosing $S_E$ in the form of
\eq{SE} and of \eq{TMIS}.

 The generating function approach given by \eq{Z}, \eq{GFEQ} and \eq{AM},
 has the advantage that it can be solved analytically (see Ref.
\cite{KOLE}). This
solution leads to a constant cross section at high energy, while the
 interaction without the four Pomeron term, decreases at high energy
\cite{AMCP}. This
fact emphasizes the importance of $s$-channel unitarity, in finding the
asymptotic
 behaviour of the scattering amplitude at high energy.  Having  an exact
solution
 we are able to develop  approximate methods, which we can check
against the
exact solution.

\subsection{Improved Mueller-Patel-Salam-Iancu approximation}
 Calculating the high energy amplitude  in the generating function approach, we
use the  approximation  which allows us to write a simple analytical formulae for the physical
observable.
The main idea of this approximation, which we call the  improved
Mueller-Patel-Salam-Iancu approximation,
is the following: we claim that at high energy, in the kinematical region
\beq \label{KR}
  Y\,\,\leq\,\,\frac{\Delta^2_\pom}{g^2_{3 \pom}}\,\,\equiv\,\,\,\,\frac{1}{\gamma},
\eeq
only large Pomeron loops, with  a  rapidity size   of the order of $Y$,
 contribute to the high energy
asymptotic behaviour of the scattering amplitudes. This approximation
 has been discussed in detail
 in Refs. \cite{MPSI,LEPR,LMP}, and here we  illustrate this idea using the
example of the first Pomeron loop
diagram given in \fig{denre}.  Using \eq{FI} or the generating
 function approach,
 we obtain   the
contribution of this diagram   in the form\footnote{In \eq{FEND} we
redefine the hadron-Pomeron vertices denoting them by
 $g(b)\,\sqrt{\gamma}$.
 We will clarify our reason for doing  so below. For
simplicity we use $g(b)$ for this vertex,
 neglecting the different hadronic states in our two channel model.}

\FIGURE[ht]{
\centerline{\epsfig{file=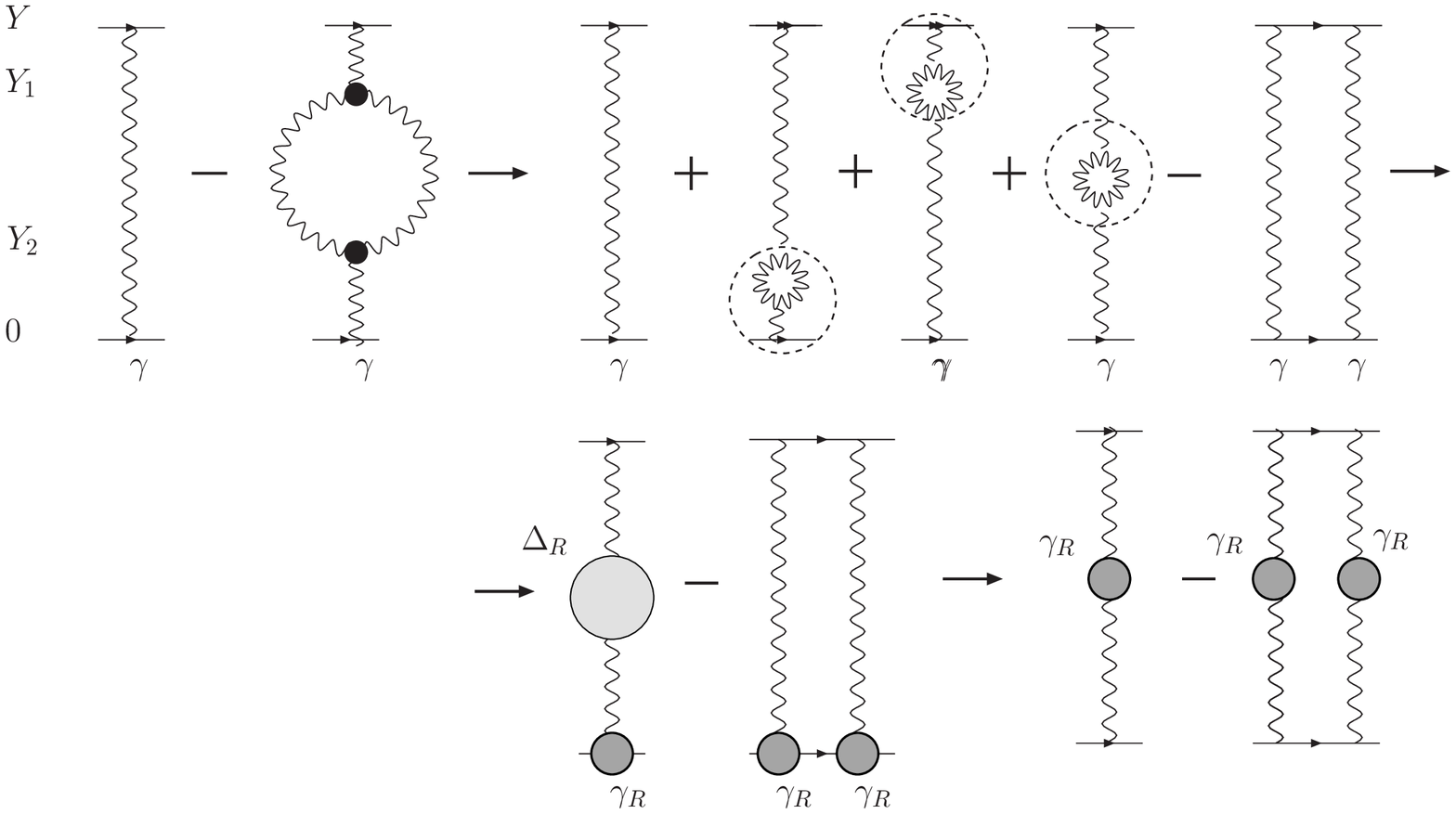,width=165mm,height=70mm}}
\caption{ The different contribution to the first enhanced diagram and its renormalization procedure .  } \label{denre} }

\bea
&&A\Lb \fig{denre}\Rb\, = \label{FEND}\\
    \,\,&&= \,g^2(b)\,\gamma\,G(Y - 0 )
\,\,-\,g^2(b)\,\Gamma(1 \to 2)\,\Gamma(2\to 1)\,\gamma\,\int^Y_0\,d\,y_1\,\int^{y_1}_0\,d\,y_2\,G(Y - y_1)\,G^2(y_1 - y_2)\,G(y_2 - 0) \notag\\
&&=\,\,\,g^2(b)\,\gamma\,G(Y - 0 )\,-\,g^2(b)\,\Gamma(1 \to 2)\,\Gamma(2\to
1)\,\gamma\,\int^Y_0\,d\,y_1\,\int^{y_1}_0\,d\,y_2\,\,e^{\Delta_\pom\,(Y
+ y_1 - y_2)}\nonumber\\
&&=\,\,g^2(b)\,\gamma\, e^{\,\Delta\_\pom\,Y}\,-\,g^2(b)\,\Gamma\Lb\,1\to\,2\Rb\,\Gamma(2 \to 1)\,\gamma\Lb
\frac{1}{\Delta^2_\pom}\,e^{2\,\Delta_\pom\,Y}\,\,-\,\,
\frac{1}{\Delta^2_\pom}\,e^{\,\Delta_\pom\,Y}\,\,-\,\,\frac{Y}{\Delta_pom}\,
      e^{\,\Delta_\pom\,Y}  \Rb  \nonumber\\
&&=\, \,g^2(b)\, \left\{ \gamma\, e^{\,\Delta_\pom\,Y}\,\,-\,\, \gamma^2 \,e^{2\,\Delta_\pom\,Y}\,\,+\,\,\gamma^2
\,e^{\Delta_\pom\,Y}\,\,+\,\,\gamma \Lb \Gamma(2\to 1)\,Y \Rb
\,e^{\Delta_pom\,Y} \right\}\notag\\
&& \longrightarrow\,\,\,\,g^2(b)\, \left\{ \gamma_R\,\,e^{\Delta_R\,Y}   \,\,-\,\,\gamma^2_R\,\,e^{ 2\,\,\Delta_\pom
\,Y}\right\}, \label{FEND1}
      \eea
with $\gamma_R\,=\,\gamma + \gamma^2$, and  the renormalized intercept of
the Pomeron $\Delta_{RE}\,=\,\Delta_\pom +\,\Gamma(2 \to 1)\,=\,\Delta_\pom \,+\,\,\Delta_\pom \,\gamma$. In the kinematic region of \eq{KR}
we can neglect the renormalization of the intercept, since in this region
$ \Delta_\pom \,\gamma\,Y\,\ll\,1$.

\begin{figure}
\begin{minipage}{11cm}
\begin{center}
 \includegraphics[width=0.80\textwidth]{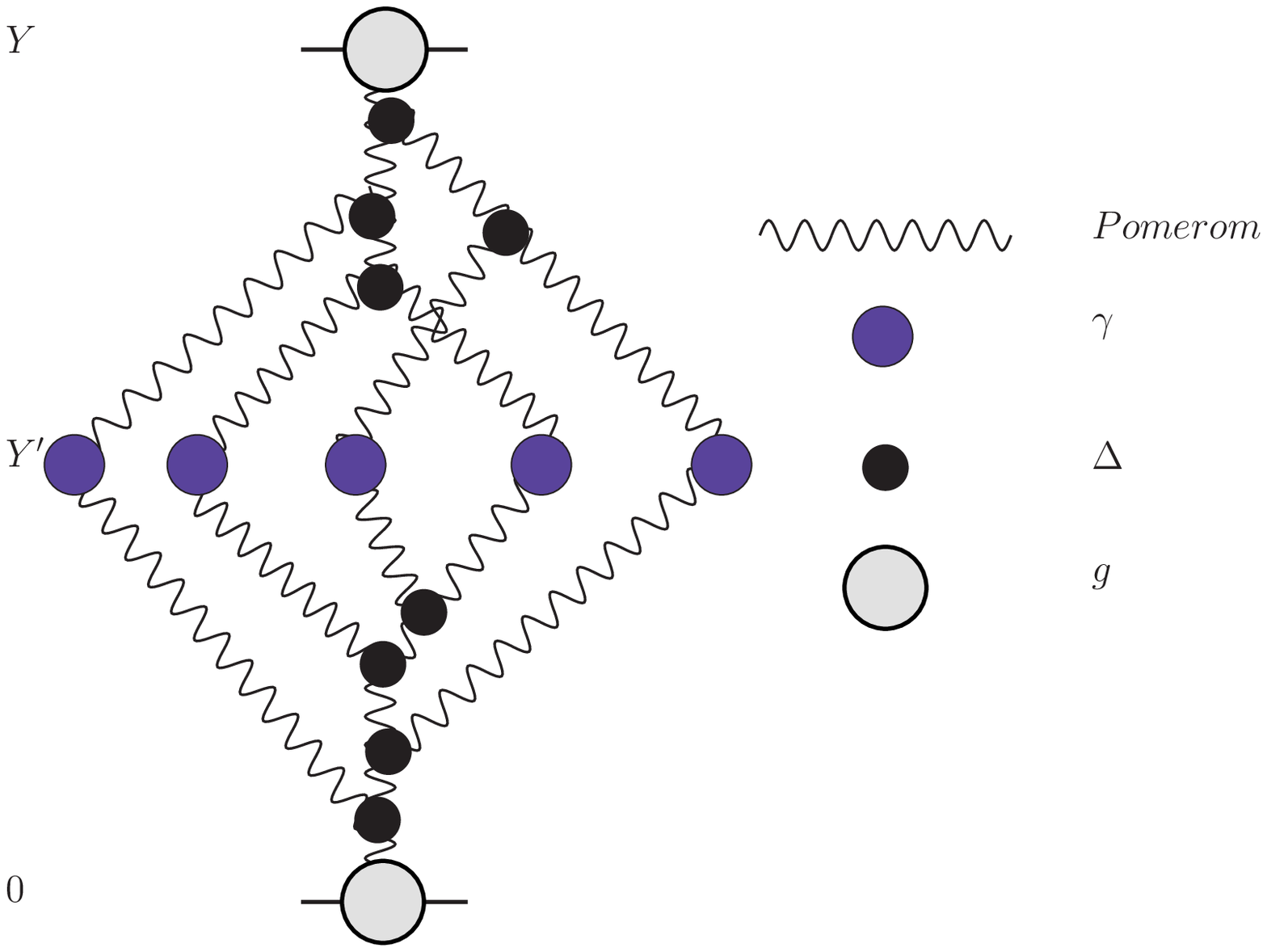}
\end{center}
\end{minipage}
 \begin{minipage}{5.5cm}
\caption{An example of enhanced diagrams, that contribute to the
unitarity constraint in the $t$-channel.  Wave  lines denote the
 Pomerons.  $\gamma$ is the
amplitude of the dipole-dipole interaction at low energies (at
rapidity $Y_0 \approx 1/\bas$ ). The particular
set of diagrams shown in this figure, corresponds to the MPSI
approach \cite{MPSI,KOLE}.}
 \label{gefunenh}
\end{minipage}
\end{figure}

The general procedure for summing large Pomeron loops was suggested by
Mueller, Patel, Salam and Iancu
(MPSI) in
Ref. \cite{MPSI}. In this approach, the scattering amplitude is calculated using the unitarity
constraints in the $t$-channel  (assuming that the amplitudes at high
energy  are purely  imaginary,
namely  $N\,=\,Im\,A$),

 \beq \label{MPSIUN}
 N([\dots]|
Y)\,\,=\,\,N([\dots]| Y - Y';P \to nP)\,\bigotimes \,N([\dots]| Y';P
\to nP). \eeq
  $\bigotimes$ stands for all necessary
integrations, while   $[\dots]$ describes all quantum numbers.
 The amplitude on the LHS  of \eq{MPSIUN},
describes all the  enhanced diagrams,  while the amplitude on the
RHS of this equation, corresponds to the splitting of  one Pomeron
to $n$ Pomerons. The precise meaning  of this equation, will become
clear in the next equation. The convenient form of \eq{MPSIUN}, has
been written \cite{GRPO,MPSI,KOLE} in terms of the generating
functional of \eq{Z}, and it takes the form

\beq \label{TUNG}
N\left(Y\right)\,\,\,=
\,\,\sum^{\infty}_{n=1}\,\,\frac{(-1)^n}{n!}\,
\,\,\gamma^n\,\,  \frac{\partial^n\,Z^p(Y -Y',\,u_p)}{\partial^n\, u_p}|_{u_p =1}
\,\,\frac{\partial^n Z^t(Y',\,u_t)}{\partial^n\, u_t}|_{u_t =1}.
\eeq
In \eq{TUNG} we denote  by $Z^p$ and $Z^t$ the generating functions that
describe projectile and target respectively.

\eq{TUNG} shows that
each dipole with rapidity $Y'$ from the target, can interact   with
any dipole from the projectile, (see \fig{gefunenh}) with the scattering amplitude  $\gamma$.
The factor $1/n!$ in \eq{TUNG} appears due to the identity of Pomerons.
\eq{TUNG}  is defined in the kinematic region of  \eq{KR}, and has a clear
physical  meaning, being  the scattering amplitude of
two partons (two dipoles)  at low energy $ Y_0\,\leq\,1/\Delta_\pom$.
\eq{TUNG} gives  a natural generalization of \eq{AM},  with an
obvious physical interpretation,  that it is the sum of terms,
and each of these terms is the product of  probabilities to find
$n$-dipoles in the projectile and target, multiplied by the
scattering amplitude.  Since we are discussing the generating
functional that satisfies \eq{INC1} as the initial condition,
\eq{TUNG} gives the sum of enhanced diagrams or, in other words, at high
energy it
leads to a new resulting Green's function of the Pomeron.

The generating functions for the projectile $Z^p \Lb Y -Y' \Rb$ and for
the  target  $Z^
t \Lb Y' \Rb$
in \eq{TUNG}, satisfy a very simple equation
that describes the parton cascades, in which a  parton can only  decay
into
two partons. This equation has the
form
\beq \label{FDEQ}
\,\,-\frac{\partial\,Z(y,\,u)}{\partial\, y}\,\,
=\,\,-\,\Delta_\pom\,u\,(1\,-\,u)
\,\,\frac{\partial\,Z(y,\,u)}{\partial\, u}.
\eeq
The above equation has the solution
\beq \label{GFSOL}
Z\Lb y, u\Rb\,\,=\,\,\frac{u}{u \,\,+\,\,(1 - u)\,e^{\Delta_\pom y}}\,\,=\,\,\frac{1}{1 \,\,+\,\,\gamma_R\,
e^{\Delta_\pom y}},
\eeq
where $u = 1/(1 + \gamma_R)$.
 \eq{GFSOL} satisfies the initial and boundary conditions of \eq{INC1} and \eq{INC2}.
Using \eq{GFSOL} and \eq{AM} it is easy to show that the amplitude is  equal to
\beq \label{GR}
N(Y)\,\,=\,\,1 - Z\Lb u = 1/\gamma_R \Rb \,\,=\,\,\sum_{n = 1}\,(-1)^n \,\, \gamma^n_{R}\,e^{ n\,\Delta_\pom
Y}.
\eeq
 \eq{GR} sums the  `fan' Pomeron diagrams and corresponds to the mean field approximation (MFA) of our
problem.
\subsection{ High energy amplitude}
Using MFA and \eq{GR}, we can rewrite \eq{TUNG} as
\bea \label{MPSI}
N^{MPSI}_{el} \left(Y\right)\,\,\,&=&
\,\,\sum^{\infty}_{n=1}\,\,(-1)^{n + 1}\frac{1}{n!}\,
\,\,\gamma^n\,\,  \frac{\partial^n\,N^{MFA}(Y -Y',\,\gamma^p_R)}{\partial^n\, \gamma^p_R}|_{\gamma^p_R =0}
\,\, \frac{\partial^n\,N^{MFA}(Y',\,\gamma^t_R)}{\partial^n\, \gamma^t_R}|_{\gamma^t_R =0}  \\
 &=& \,\,1\,\,\,-\,\,\,\left\{\exp \Lb
-\,\gamma\,\frac{\partial}{\partial \,\gamma^p_{R}}\,
\frac{\partial}{\partial \,\gamma^t_{R }}\Rb\,\,\,
N^{MFA}\Lb Y  - Y',\, \gamma^p_{R}\Rb\,\,N^{MFA}_0\Lb Y',  \gamma^t_{R}\Rb
\,\right\}\mid_{\gamma^p_{R} =0; \,\gamma^t_{R}  = 0}. \notag
\eea
Substituting \eq{GFSOL} into \eq{MPSI}, we obtain that \cite{KODD,KKLM}
\beq \label{AMMPSI}
N^{MPSI}_{el} \left(Y\right)\,\,\,=\,\,\,1\,\,-\,\,\exp \Lb \frac{1}{T(Y)}\Rb\,\frac{1}{T(Y)}\,\,
\Gamma\Lb 0,\frac{1}{T(Y)} \Rb,
\eeq
where $\Gamma \Lb 0,x \Rb$ is the incomplete gamma function  (see  formsph1.epsulae {\bf
8.350 - 8.359} in Ref. \cite{RY}) and
\beq \label{T}
T\Lb Y\Rb\,\,\,=\,\,\gamma\,e^{\Delta_\pom\,Y}.
\eeq
In Ref. \cite{KKLM}, the solution given by \eq{AMMPSI},  is compared with
 the exact solution (see Fig.13 of Ref. \cite{KKLM}). It turns out to
 within a 5\% accuracy, that the MPSI approximation describes the high energy
behaviour of the amplitude.  At high energy \eq{AMMPSI} gives $N^{MPSI}_{el} \left(Y\right)\,\,\to\,\,1$.

As has been mentioned, since \eq{TUNG} satisfies the initial condition of
 \eq{INC1}, it describes, as well,
\eq{AMMPSI},
 the set of enhanced diagrams, and gives the resulting Pomeron
Green's function
\beq \label{GRENF}
G^{MPSI}_\pom\Lb Y - Y'\Rb \,\,\,=\,\,\,N^{MPSI}_{el} \left(Y - Y'\right).
\eeq
Replacing the bare Pomeron Green's function by \eq{GRENF}, we can use
\eq{EL},  \eq{XST} and \eq{XSEL}, to calculate
elastic and total cross sections.
\subsection{Diffractive production processes}
\begin{figure}[h]
\begin{minipage}{11cm}
\begin{center}
 \includegraphics[width=0.99\textwidth]{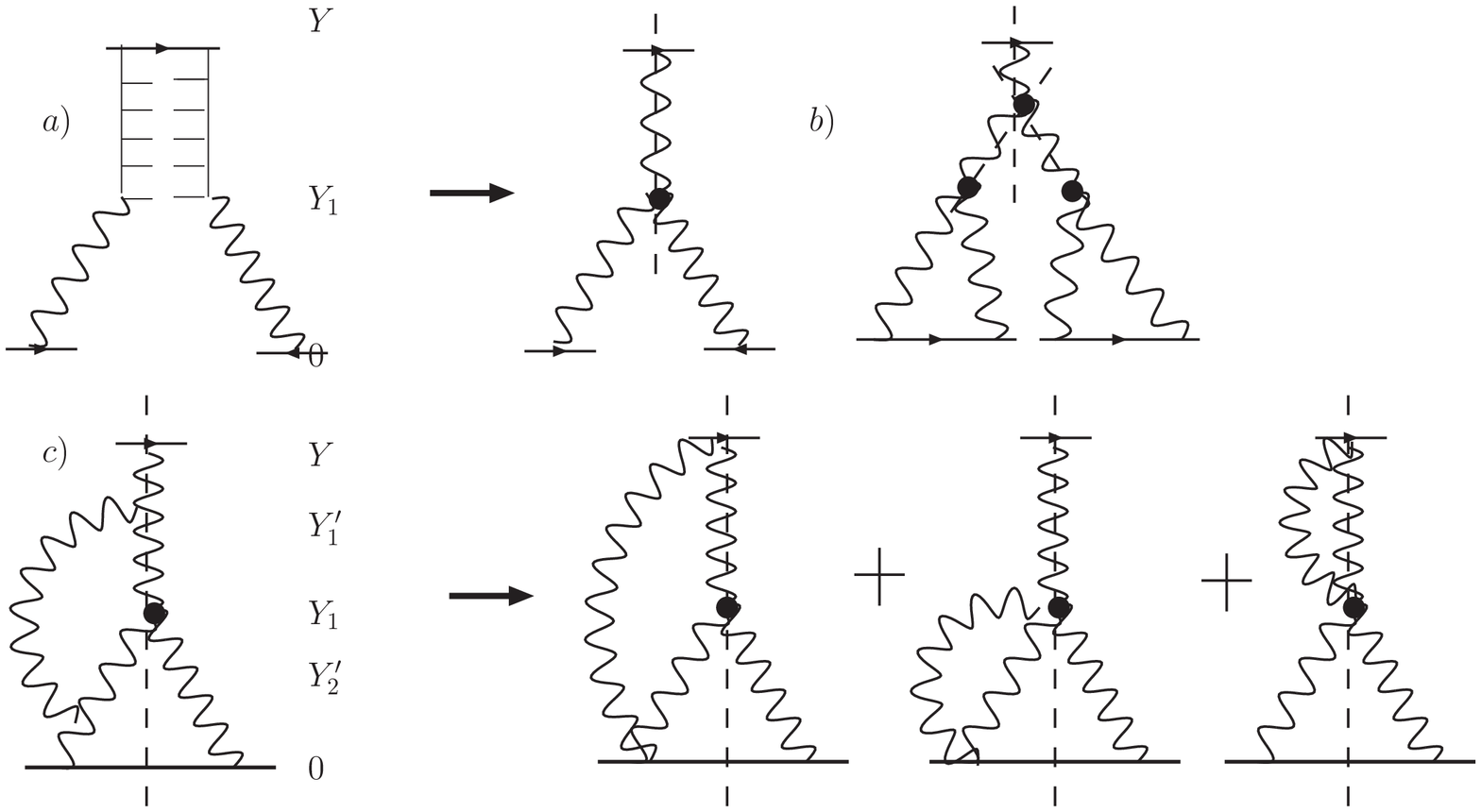}
\end{center}
\end{minipage}
 \begin{minipage}{6cm}
\caption{Several examples of the Pomeron diagrams that lead to a different source of the diffractive dissociation
 that cannot be described in the framework of the G-W
mechanism. \protect\fig{sdex}-a is the
simplest diagram that describes the  process of diffraction in the region of large mass $Y - Y_1 = \ln(M^2/s_0)
$. \protect\fig{sdex}-b and \protect\fig{sdex}-c  give  examples  of more complicated diagrams in the
region ofsph1.eps
large mass. The dashed line shows the cut Pomeron, which describes the
 production of hadrons (see \protect\fig{sdex}
-a which illustrates this point). }
 \label{sdex}
\end{minipage}
\end{figure}

The conclusion derived from the previous discussion, is that
we cannot use the formulae of \eq{SD} and \eq{DD}, for  calculating
the single
 and double diffraction cross
sections. Indeed, these two equations describe  the diffraction production due
to the G-W mechanism, while the
sum of all enhanced diagrams  leads to a new source of  diffractive production (see \fig{sdex} for examples
of such processes).

\begin{figure}[h]
\begin{minipage}{11cm}
\begin{center}
 \includegraphics[width=0.70\textwidth]{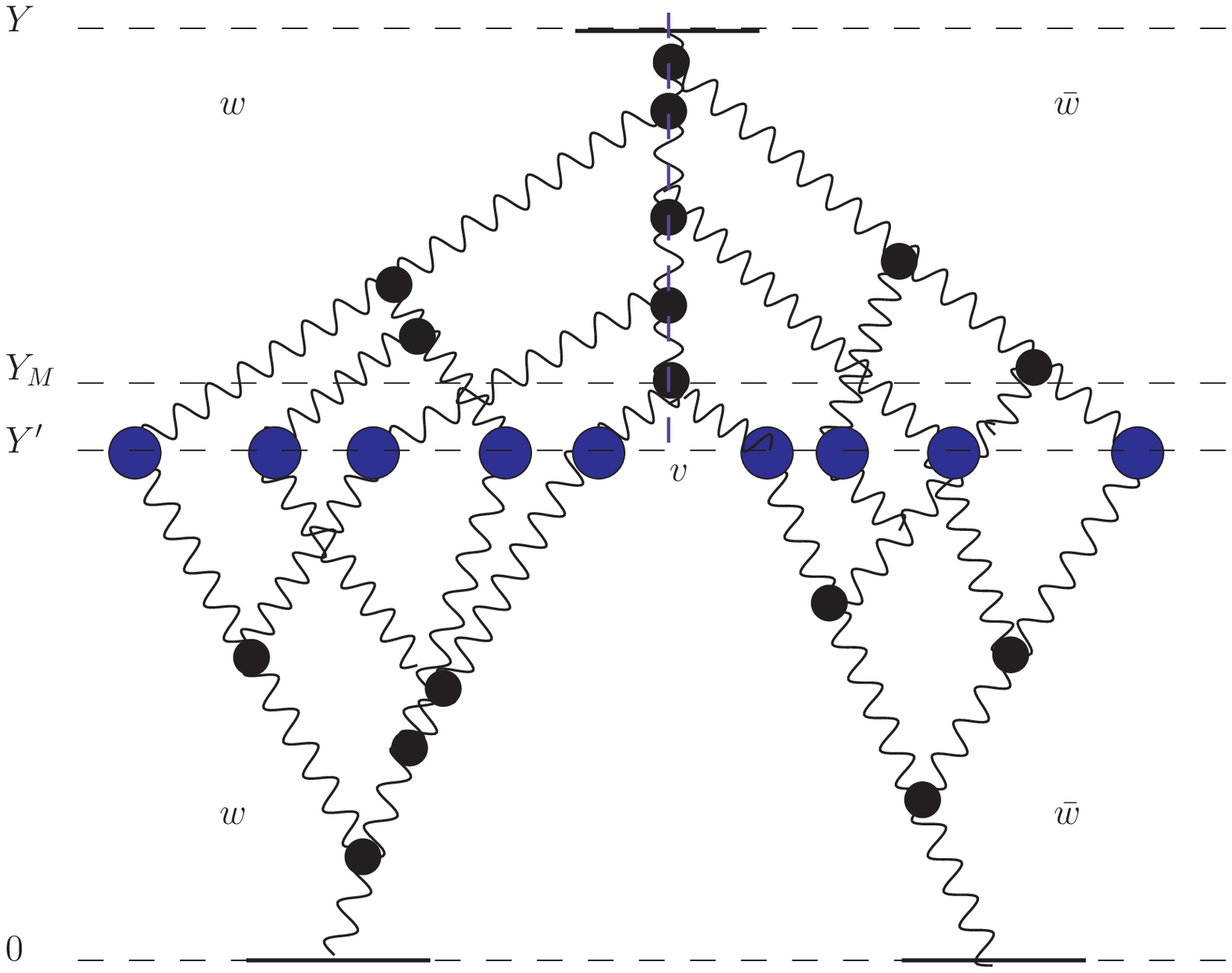}
\end{center}
\end{minipage}
 \begin{minipage}{6cm}
\caption{The MPSI approximation for the cross section of  single diffractive production of  mass ($M^2$,$
Y - Y_M = \ln (M^2/s_0)$). The dashed lines shows the cut Pomerons.  All other notations, are  as in
\protect\fig{gefunenh}.
 }
 \label{sdmpsi}
\end{minipage}
\end{figure}

We use the MPSI approximation to obtain the expression for the  additional contribution to G-W
mechanism. The main idea is shown in \fig{sdmpsi}. As was discussed  in Ref. \cite{LEPR} , in order to
apply the
MPSI approach to diffractive production, we need to consider the
generating function of three variables: $w$,$\bar{w}$
 and $v_{in}$,
\beq \label{ZINEL}
Z\Lb w,\bar{w}; v_{in};Y \Rb \\,\,\,=\,\,\sum^{\infty}_{n=0; m=0; k=0} \,P\Lb n,n,k|Y\Rb
\,\,w^n\,\bar{w}^m\,v^k_{in}.
\eeq
 $P\Lb n,n,k|Y\Rb$ is the probability to find $n$  and $m$  Pomerons in the amplitude and conjugated
amplitude  respectively, while $k$ is the number of cut Pomerons (see
Ref. \cite{AGK}).
  $Z\Lb w,\bar{w}; v_{in};Y \Rb$ has been found in Ref. \cite{LEPR} for
the parton cascade, with a decay
of one parton to two partons,
\bea \label{ZMFAD}sph1.eps
&&Z\Lb  w,\bar{w},v_{in} | Y - Y_M = \ln(M^2/s_0) \equiv Y_m \Rb\,\,=\,\, \\
&&\frac{ w \, e^{-\Delta_\pom   Y}}{1+w(e^{- \Delta_\pom  Y_m}-1)}\,\,+\,\,
\frac{\bar{w} \, e^{-\Delta_\pom   Y}_m}{1+\bar{w} (e^{- \Delta_\pom  Y_m}-1)}\,\,
  -\,\,\frac{(w + \bar{w} - v_{in}) e^{- \Delta_\pom   Y_m}}{1+(w + \bar{w}\,-v_{in})(
e^{- \Delta_\pom\,\, Y_m}-1)}. \nonumber
\eea

 To find the cross section for  single diffractive production,  we need to
calculate the term
which is
proportional to $v_{in}$, and to replace $v_{in}$ by $2 \Delta_\pom w
\bar{w}$. Indeed,  this term means that
at  $Y = Y_m$ we have only one cut Pomeron, while all other cut Pomerons
at this rapidity have decayed
to Pomerons without cuts. For simplicity  we choose $Y' = Y - Y_M = Y_m$
 (see \fig{sdmpsi}).
Replacing $  v_{in} \,\to\,2 \Delta_\pom w \bar{w}$,  means that  at $Y = Y_M$,
the last cut Pomeron splits
into two Pomerons (see \fig{sdmpsi}). Therefore,
\beq
 N^{MFA}_{sd}\Lb  w,\bar{w} ; Y - Y_M = \ln(M^2/s_0) \equiv Y_m \Rb\,\,=\,\,2\,\Delta_\pom w\,\bar{w}\,
\frac{e^{\Delta_\pom Y_m}}{\Lb 1\,\,+\,\,(w \,+\,\bar{w})\,(e^{\Delta_\pom Y_m} \,-\,1)^2 \Rb^2}.
\eeq

The general \eq{TUNG} can be rewritten in this case in the form
\bea
&&N^{MPSI}_{sd}(Y,Y_m = \ln (M^2/s_0))\,\,=\label{SD1} \\
&&\,\,\sum^{\infty}_{n=1;m=1}\,\frac{( - 1)^{n + m}}{\,n!\,m! }\,\gamma^{n + m}
 \frac{\partial^n\, }{\partial^n\, w^p}
 \frac{\partial^m\,}{\partial^n\, \bar{w}^p}\,\,N^{MFA}_{sd}\Lb  w^p,\bar{w}^p ; Y - Y_M = \ln(M^2/s_0)
  \equiv Y_m \Rb |_{w =1; \bar{w}=1} \,\notag\\
&&\times\,\,
 \frac{\partial^n\,N^{MFA}\Lb w^t,Y - Y_m \Rb }{\partial^n\, w^t}|_{w^t=1}
 \frac{\partial^m\,N^{MFA}\Lb \bar{w}^t,Y - Y_m \Rb}{\partial^n\, \bar{w}^t}|_{\bar{w}=1}
\notag\\
&& = \,1 \,-\,\left\{ \exp\Lb - \gamma \Lb
\frac{\partial\, }{\partial\, w^p}\,\frac{\partial\, }{\partial\, w^t}\,\,+\,\,\frac{\partial\, }{\partial\, \bar{w}^p}\,\frac{\partial\, }{\partial\, \bar{w}^t} \Rb \Rb
\,
N^{MFA}_{sd}\Lb  w,\bar{w} ; Y_m \Rb\, \right.
\notag \\
&& \left.\times \,\,N^{MFA}\Lb \bar{w}^t,Y - Y_m = Y_M\Rb\,N^{MFA}\Lb \bar{w}^t,Y - Y_m = Y_M\Rb \right\}|_{w^p = w^t=\bar{w}^p =\bar{w}^t = 1}\notag \\
&&= \frac{\Delta_\pom \gamma^2}{6}\,\frac{e^{ \Delta_\pom ( 2\,Y - Y_m)}}{L^2\Lb Y,Y_m\Rb}\,G\Lb L\Lb Y,Y_m\Rb\Rb,
\label{SD2}
\eea
where
\beq \label{GL}
  G\Lb L\Rb\,\,=\,\,L\,( (L-1)^2 -2 ) + e^{1/L}( 1 + 3L)\,\Gamma_0(1/L)
  \eeq
and
\beq \label{L}
 L \Lb Y,Y_m \Rb\,\,=\,\, \gamma\,\exp\Lb \Delta_\pom (Y - Y_m)\Rb\,\left\{ \exp\Lb \Delta_\pom Y_m\Rb - 1 \right\}.
\eeq
$N^{MPSI}_{sd}(Y,Y_m = \ln (M^2/s_0))$ describes the differential single diffraction cross section for
the production of mass $M$.  We can calculate the integrated  diffraction cross section
\bea \label{TSD}
N^{MPSI}_{diff}\Lb Y; M_{max},M_{min} \Rb \,\,&\equiv &\,\,\int^{y_m(max) }_{y_m(min) }\!\!\!\!\!\!\!\! d \,y_m \,N^{MPSI}_{sd}(Y,Y_m = \ln (M^2/s_0))\,\,\notag \\
&=&\,\,\frac{\gamma}{6}\,e^{\Delta_\pom Y}\,\Lb
 B\Lb L\Lb Y,Y_m(max)\Rb\Rb\,\,-\,\, B\Lb L\Lb Y,Y_m(min)\Rb\Rb\Rb ,
\eea
where $Y_m(max)= \ln(M^2_{max}/s_0)$ , $Y_m(min)= \ln(M^2_{min}/s_0)$ and
\beq \label{BL}
B\Lb L\Rb\,\,=\,\, 2 \,+\,\frac{1}{L^2} \,-\,\frac{1}{L}\,\,-\,\,\frac{e^{\frac{1}{L}}\Gamma_0\Lb
1/L\Rb}{L^3}.
\eeq
$M_{max}$ and $M_{min}$ denote the largest and the smallest masses  which
are produced in  the diffractive
process.

\begin{figure}[h]
\begin{minipage}{11cm}
\begin{center}
 \includegraphics[width=0.99\textwidth]{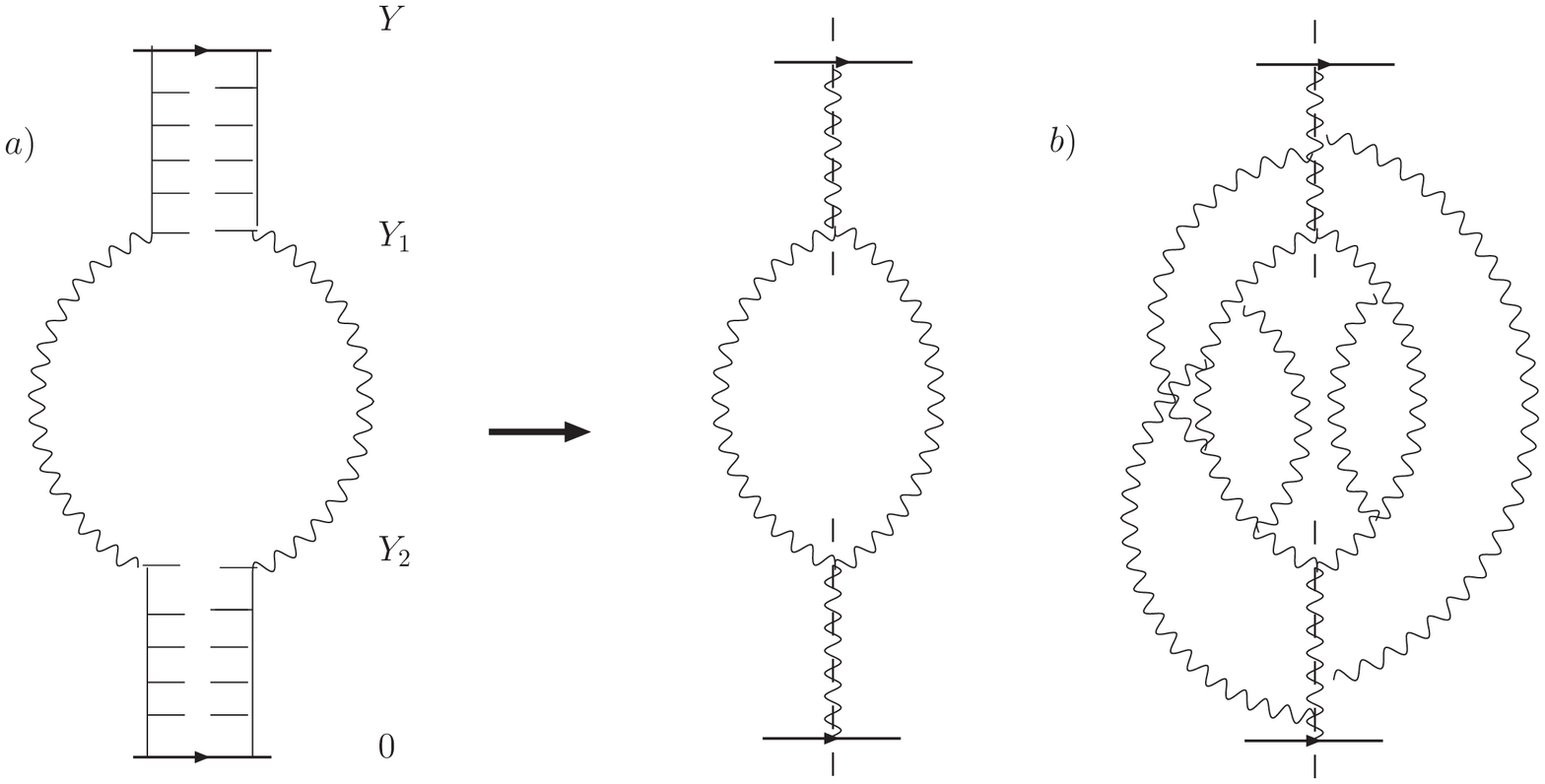}
\end{center}
\end{minipage}
 \begin{minipage}{6.5cm}
\caption{Several examples of the Pomeron diagrams that lead to
the double diffractive production.
\protect\fig{ddex}-a is the
simplest diagram that describes the  process of  double  diffraction in
the regions of large mass $Y - Y_1 = \ln(M^2_1/s_0)$ and
 $Y_2  = \ln(M^2_2/s_0)$.
  \protect \fig{sdex}-b  contains  examples  of more complicated diagrams
in the
region of large masses. The dashed line indicates the cut Pomeron which
describes the
production of hadrons (see \protect\fig{ddex}-a). }
 \label{ddex}
\end{minipage}
\end{figure}

The expression for the integrated cross section for  double diffraction  can
be
 obtained directly from the unitarity constraint of \eq{UNIT}, as
the
diagrams that describe the elastic and single diffraction cross sections,
do
not contribute to  the set of Pomeron diagrams, that describe the exact
Pomeron Green's function (see \fig{ddex} which contains  examples of the
diagrams that contribute to  double diffractive production).

The unitarity constraint is given by
\beq \label{UNDD}
2 \,N^{MPSI}\,\,=\,\,N^{MPSI}_{dd} \,\,+\,\,N^{MPSI}_{in},
\eeq
where $N^{MPSI}_{in}$ stands for the inelastic cross section.  It was
 shown that $N^{MPSI}_{in}$ is equal to $
N^{MPSI}\Lb 2 T(y)\Rb$ (see Refs. \cite{KL,BORY,LEPR}).  Therefore, the integrated double diffraction
cross section can be written in the form
\beq \label{TDD}
N^{MPSI}_{dd} \Lb Y\Rb\,\,\,=\,\,\,2 \,N^{MPSI} \Lb T( Y)\Rb \,\,\,-\,\,\,N^{MPSI} \Lb 2\,T( Y)\Rb.
\eeq
\subsection{Our approach}
In our approach we combine the G-W mechanism with the exact Pomeron
Green's
function of \eq{GRENF}. First, we replace the bare Pomeron Green's
function
 $G(Y - Y') \,=\,\,\exp\Lb - \Delta_\pom \,( Y - Y')\Rb$ in \eq{OMEGA} by
 $ G^{MPSI}_\pom\Lb Y - Y'\Rb $ of \eq{GRENF}, and, obtain
\beq \label{OMEGAF}
\Omega_{i,k}\Lb s,b \Rb\,=\,g_{i} \,g_{k}\,\,G^{MPSI}_\pom\Lb Y - Y'\Rb \,\,
\,S\Lb b ;m_i,m_k;  \alpha'_\pom \ln (s/s_0) =0 \Rb ,
\eeq
with a profile function $\,S\Lb b ;m_i,m_k;  \alpha'_\pom \ln (s/s_0) \Rb$ determined by \eq{SDB}.

Using \eq{EL},  \eq{XST} and \eq{XSEL}  with $\Omega_{i,k}\Lb s,b \Rb$ from \eq
{OMEGAF}, we can calculate
elastic and total cross sections. These formulae provide a correct description
 of  low mass diffractive production due to the  G-W  mechanism. However, to
include the Pomeron interactions in processes of diffractive production,
 we need to change  \eq{SD} and \eq{DD}.  We first introduce
 (see \fig{sdex}-a)
\beq \label{OMEGASD}
\Omega^{sd}_{i,k}\Lb s,M;b\Rb\,\,\,=\,\,g_i\,g^2_k\,N^{MPS}_{sd}\Lb Y\,=\,\ln(s/s_0),Y_m =\ln(M^2/s_0)\Rb
\,S^{sd}\Lb b; m_i,m_k\Rb,
\eeq
where the new profile function $S^{sd}\Lb b; m_i,m_k\Rb$  is the Fourier
 transform of
\beq \label{SSDB}
\frac{1}{1 +  q^2/m^2_i)^2}\,\int\frac{ d^2 k}{(2 \pi)^2}\frac{1}{1 + ( \vec{q}  + \vec{k})^2/m^2_i)^2}\,
\frac{1}{1 + ( \vec{q}  - \vec{k})^2/m^2_i)^2}\,\,\Longrightarrow \,\,\,S^{sd}\Lb b; m_i,m_k\Rb.
\eeq
We found  to within  an accuracy of around 5\%,that  the profile function
$S^{sd}\Lb b; m_i,m_k\Rb$
can be approximated by
\beq \label{SSD}
S^{sd}\Lb b; m_i,m_k\Rb\,\,\,=\,\,\,\frac{ m^2_k}{12\, \pi}\,S\Lb b ;m_i,\bar{m}_k;  \alpha'_\pom \ln (s/s_0) =
0\Rb\,\,\,\,\, ,
\eeq
where
$\bar{m}^2_k\,\,=\,\,2\,\sqrt{3}\,m^2_k$ .

For the calculation of  the integrated cross section of   single
diffractive production we
 define
\bea \label{OMEGATSD}
&&\Omega^{sd}_{i,k}\Lb s,M_{max}, M_{min};b\Rb\,\,\,= \\
&&\,\,g_i\,g^2_k\,N^{MPS}_{diff}\Lb Y\,=\,\ln(s/s_0),Y_m(max) =\ln(M^2_{max}/s_0), Y_m(min) =\ln(M^2_{min}/s_0)\Rb
\,S^{sd}\Lb b; m_i,m_k\Rb.\notag
\eea
For the integrated cross section of the single diffraction channel, we
obtain the following
expression which takes into account both the  G-W  mechanism, and the
enhanced Pomeron diagrams:
\bea
\sigma_{diff} = \int^{M^2_{max}}_{M^2_{min}}\!\!\!\frac{d M^2}{M^2}\,
\frac{d \sigma^{sd}(s,M)}{d M^2}
 & = &\int d^2 b  \left\{\,\,\alpha^2\,\beta^2\Lb \alpha^2\, e^{ - \frac{\Omega_{1,1}}{2}}\,-\,( \alpha^2 - \beta^2 )\,e^{ - \frac{\Omega_{1,2}}{2}}\,-\,\beta^2\,e^{ - \frac{\Omega_{2,2}}{2}}\Rb^2 \right. \label{DI1}\\
 &+& \left. \left\{ \alpha^2\,\Omega^{diff}_{1,1}
 e^{ - \Omega_{1,1}}\,+\,\beta^2\,\Omega^{diff}_{2,2}\, e^{ - \Omega_{2,2}}\,
\,+\,2\,\alpha^2\,\beta^2\,\Omega^{diff}_{1,2} e^{ - \Omega_{1,2}}\right\} \right\}, \label{DI2}
\eea
 where $\Omega^{diff}_{i,k}$ are given by \eq{OMEGATSD}.
The differential cross section for single diffraction in the region
$ M \;\;>\;\; M_{min}$ is
\beq \label{DXSSD}
  M^2\,\frac{d \sigma^{sd}(s,M)}{d M^2}\,=\,\int d^2 b \left\{ \alpha^2\,\Omega^{sd}_{1,1}
 e^{ - \Omega_{1,1}}\,+\,\beta^2\,\Omega^{sd}_{2,2}\, e^{ - \Omega_{2,2}}\,
\,+\,2\,\alpha^2\,\beta^2\,\Omega^{sd}_{1,2} e^{ - \Omega_{1,2}}\right\},
\eeq
where $ \Omega^{sd}_{i,k}$ are given by \eq{OMEGASD}.
In \eq{DI1} and \eq{DXSSD} we use $\Omega_{i,k}$ of \eq{OMEGAF}.

For the double diffractive cross section we introduce
\beq  \label{OMEGADD}
\Omega^{dd}_{i,k}\Lb s,b \Rb\,=\,g_{i} \,g_{k}\,\,N^{MPSI}_{dd} \Lb Y \Rb \,\,
\,S\Lb b ;m_i,m_k;  \alpha'_\pom \ln (s/s_0)=0 \Rb .
\eeq
Using this $\Omega^{dd}_{i,k}$ we obtain for the integrated DD cross section
\bea
\sigma_{dd} &=& \int^{\infty}_0\!\!\!\frac{d M^2_1}{M^2_1}\,\int^{\infty}_0\!\!\!\frac{d M^2_2}{M^2_2}
\frac{d \sigma^{sd}(s,M_1,M_2)}{d M^2_1\,d M^2_2}
 = \int d^2 b  \left\{\,\,\alpha^4\,\beta^4\Lb  e^{ - \frac{\Omega_{1,1}}{2}}\,-\,2\,\,e^{ - \frac{\Omega_{1,2}}{2}}\,+\, \,e^{ - \frac{\Omega_{2,2}}{2}}\Rb^2 \right. \label{DIDD1}\\
 &+&\left. \,\left\{ \alpha^4\,\Omega^{dd}_{1,1}
 e^{ - \Omega_{1,1}}\,+\,\beta^4\,\Omega^{dd}_{2,2}\, e^{ - \Omega_{2,2}}\,
\,+\,2\,\alpha^2\,\beta^2\,\Omega^{dd}_{1,2} e^{ - \Omega_{1,2}}\right\} \right\}. \label{DID2}
\eea
  For  the integrated single and double diffractive production, the expressions
each contains two terms: the first  is responsible for
G-W mechanism for these processes, while the second
  originates from the large mass diffraction of the enhanced Pomeron diagrams.

 \eq{EL},  \eq{XST} and \eq{XSEL}  with $\Omega_{i,k}\Lb s,b \Rb$ from \eq{OMEGAF}
and \eq{OMEGASD} - \eq{DID2} give the full list of formulae in our
approach.
We wish to emphasize that our approach  is based on the pQCD
 input for calculating the enhanced Pomeron diagrams, and on the G-W
 mechanism for low mass diffractive production.

\section{Results of the fit}
\subsection{Cross sections and elastic slope}
  We have adjusted the  parameters of our model
 which  are listed in Table 2, using the formulae of section 3.6.
The fit is  based on
 55 experimental data points,
which includes the $p$-$p$ and $\bar{p}$-$p$ total cross sections,
integrated elastic cross sections,
integrated single and double  diffraction cross sections,
and the forward slope of the elastic cross section
in the ISR-Tevatron energy range. The model gives
 a good reproduction of the
data,  with a $\chi^2/d.o.f. \,\approx 1.25$.
The quality of description of the experimental data is shown in \fig{xst},
\fig{bel}, \fig{xsel}, \fig{xssd} and \fig{xsdd}.  A significant
contribution
to $\chi^2/d.o.f. $ stems from the uncertainty for the value of two single
diffraction cross sections, and
 of the total cross section
 at the Tevatron.  The $\chi^2/d.o.f.$ in  Table 2 is calculated
neglecting the contribution of the CDF measurement [41]
($\sigma_{tot}$ = 80
mb. at the Tevatron energy).         The important advantage
of our approach,  is that
the model provides a very good reproduction of the DD data points. In our
previous
attempt to describe the DD data
 \cite{GLMLAST} within a G-W approach,  it was necessary to  assume
 a non-factorizable contribution
for the Pomeron exchange, resulting in marginally acceptable results.

\DOUBLEFIGURE[ht]{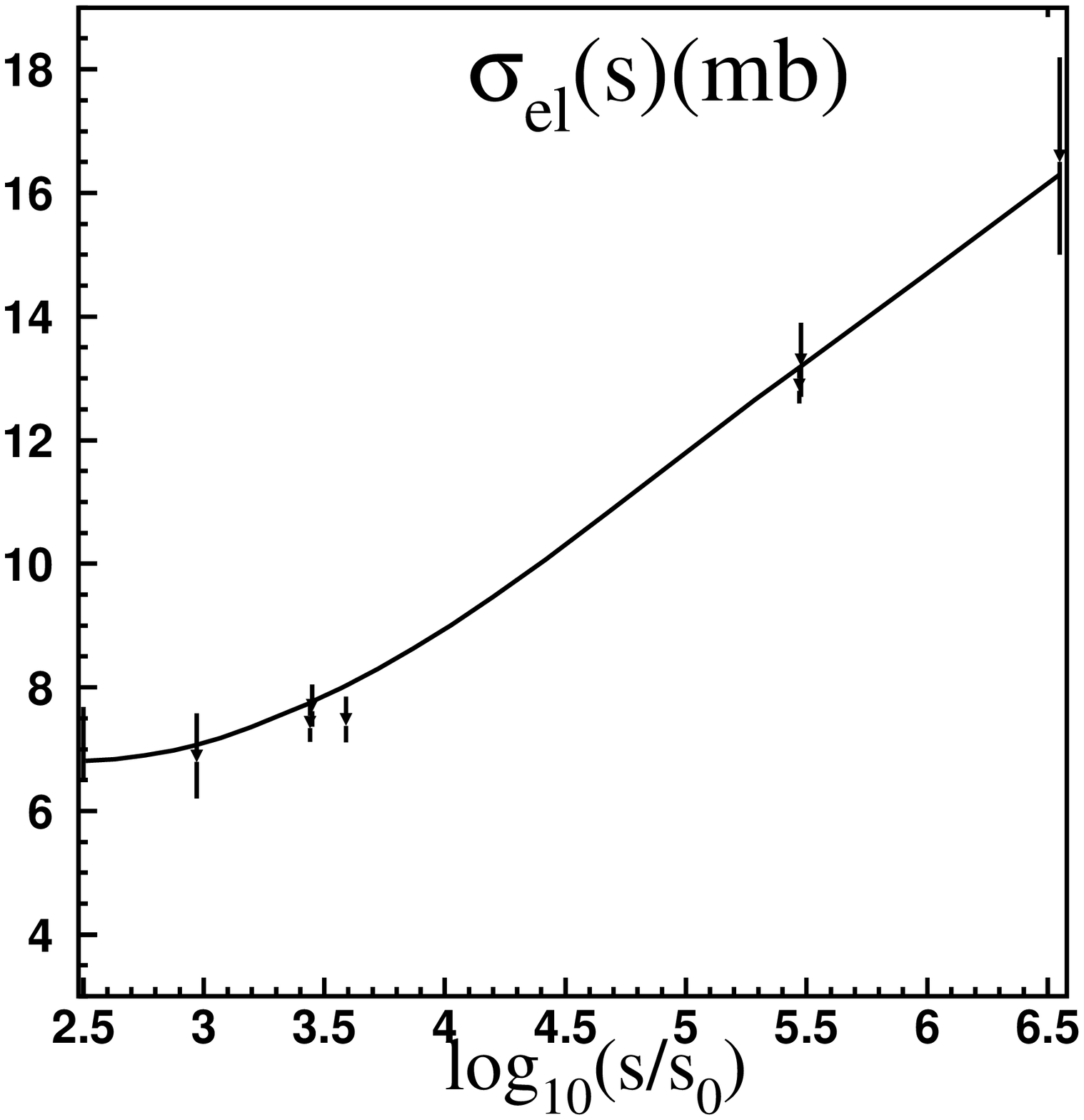,width=85mm,height=70mm}{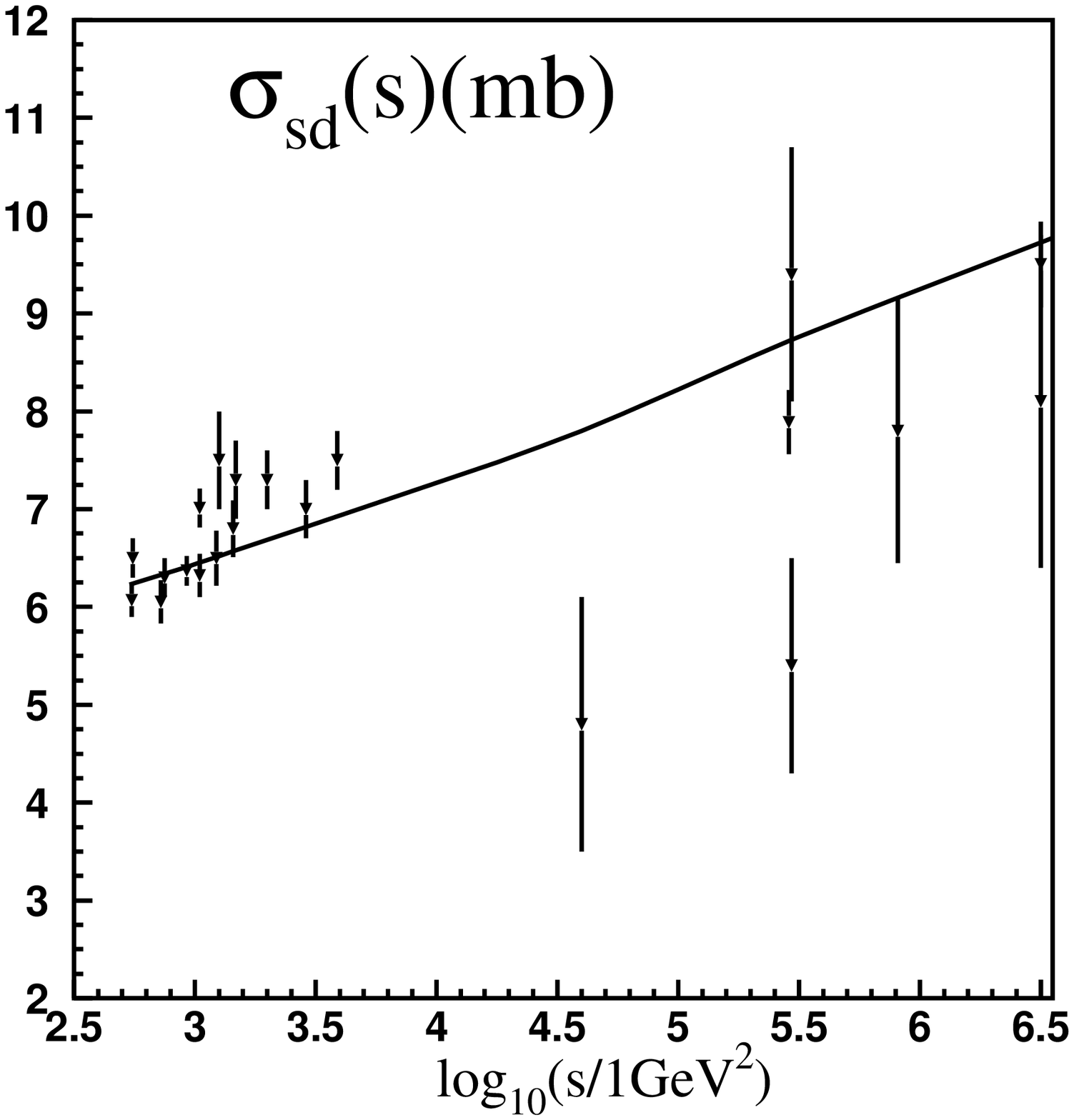,width=85mm,height=70mm}
{Energy dependence of $\sigma_{el} $.
\label{xsel}}
{Energy dependence of the cross section $\sigma_{sd}$.
\label{xssd}}

\DOUBLEFIGURE[ht]{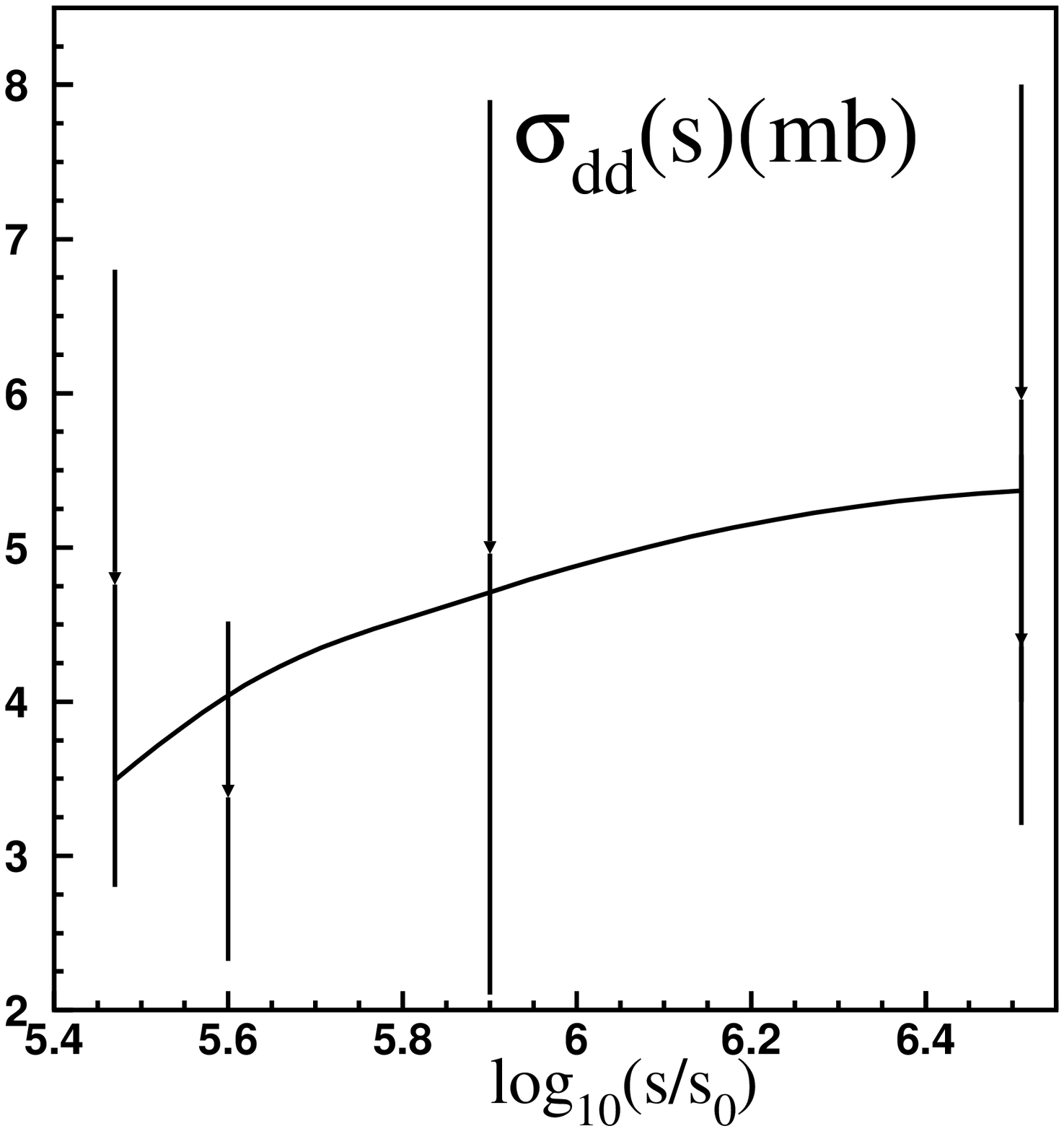,width=85mm,height=70mm}{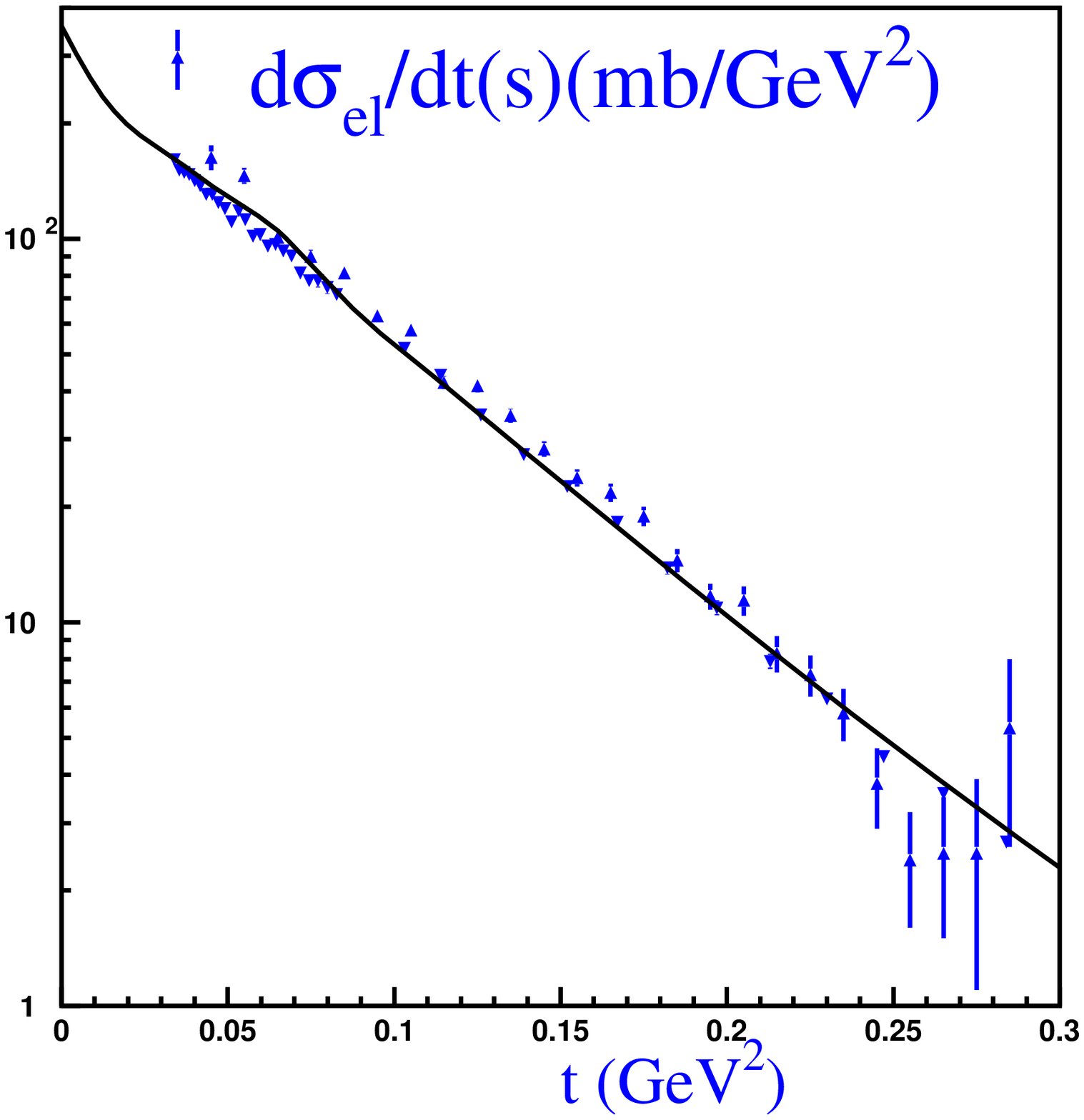,width=85mm,height=70mm}
{Energy dependence of  $\sigma_{dd }$.
\label{xsdd}}
{$t$ dependence of  $d \sigma_{el}/d t$ at the Tevatron ($W = 1800\,GeV$).
\label{dxsdt}}

\TABLE[ht]{
\begin{tabular}{|l|l|l|l|l|l|l|}
\hline
 $ \Delta_\pom $ & $\beta$ &  $\alpha^{\prime}_{\pom}$& $g_1$ &  $g_2$ & $m_1$ &
$m_2$  \\ \hline
0.335 & 0.339 & 0.012 $GeV^{-2}$ &5.82(0.90) $GeV^{-1}$ &
239.6(37.27) $GeV^{-1}$ & 1.54 $GeV$& 3.06 $GeV$
\\ \hline
$ \Delta_{\reg} $ & $\gamma $ &  $\alpha^{\prime}_{\reg}$& $g^{\reg}_1$ &  $g^{\reg}_2$ & $ R^2_{0,1}$& $\chi^2/d.o.f.$
 \\ \hline
-\,0.60 & 0.0242 & 0.6 $GeV^{-2}$ & 13.22 $GeV^{-1}$ & 367.8 $GeV^{-1}$ & 4.0 $GeV^{-2}$ &
1.0
\\ \hline
\end{tabular}
\caption{Fitted parameters for our model which includes G-W mechanism for
diffractive production, as well as the diffractive processes that stem
from enhanced Pomeron
 diagrams (Pomeron loops). In our approach we have  extracted  a factor of
$\gamma$
 from the product $g_i g_k$.
 Therefore, the value of the Pomeron-hadron vertex
 is equal to $g_i = g_i(\mbox{from the table)} \times \sqrt{\gamma}$,
and
this value  is shown in parentheses. }
\label{t2}}

{\boldmath
\subsection{$t$ - dependence of the differential elastic cross section}
}

The behaviour of the elastic cross section in the region of small $t$, is
characterized by $B_{el}$ and $\sigma_{tot}$,
which are included
in the set of experimental data that
 we use for our fit.  However, we wish  to know the scattering amplitude at
a
relatively high value of $t \geq
0.1\, GeV^2$,  where the simple exponential $t$  behaviour of the input
elastic amplitude
 does not describe the data \cite{GLMLAST}.
  In \fig{dxsdt} we plot our prediction  with the parameters of Table 2 for the
 $t$-behaviour
 of the elastic cross section at the Tevatron energy $W = 1800\,GeV$.  We reproduce the data quite
well,  and it is a considerable improvement over the results  obtained in
\cite{GLMLAST}.
\footnote{We
thank M. Ryskin who pointed out that the model of Ref. \cite{GLMLAST},
gives  a minimum at $ |t| \approx 0.1 \,GeV^2$
which contradicts the experimental data.}

\subsection{Mass dependence of the diffractive cross section}
\begin{figure}[h]
\begin{minipage}{11cm}
\begin{center}
 \includegraphics[width=0.80\textwidth]{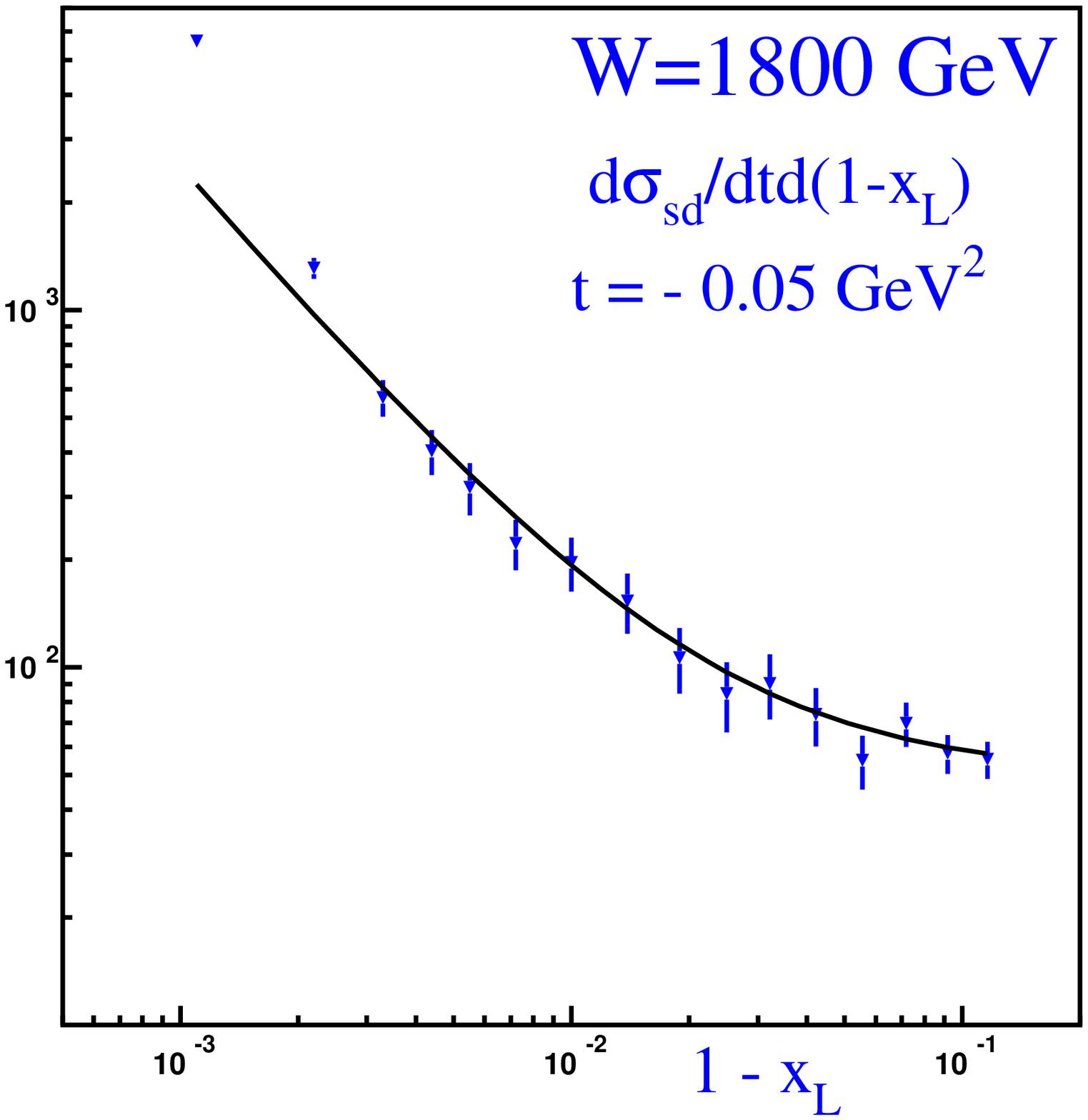}
\end{center}
\end{minipage}
 \begin{minipage}{6cm}
\caption{Dependence of single inclusive cross section on $1 - x_L =
M^2/s$,
 where $M$ is the mass of the diffractively produced system. Data are
taken from
 Refs. \cite{GOMO,CDFDD}.  }
 \label{dsdm}
\end{minipage}
\end{figure}

In \fig{dsdm}, we plot  the cross section of single diffraction as a
function of
 $1 - x_l = M^2/s$ at the Tevatron energy.  For this cross section, in the
region of
 high mass, we use \eq{DXSSD},  while for  diffraction in the region of
low mass, we
 need to make some assumptions regarding the
 dependence of this cross section on
$M$. Following
 Refs. \cite{CDFDD,KMRNEW} we assume that the main contribution for the
G-W part
 of single diffractive production stems, from the $\reg\pom \reg$ term, which
does
 not depend on $x_L$. Therefore, the resulting contribution   has the following
 form:
\bea \label{DSDM}
&&M^2 \frac{d \sigma_{sd}\Lb s,M \Rb}{d M^2\,d t}=\\
&&\,\,\,\,\,\,\,\,\,\,\,\,\,\,\,\,\,\,\,\,\,\,\,\,\,\,\,\,\,\,=\,M^2
\frac{d \sigma
^{\mbox{Low M}}_{sd}\Lb s,M \Rb}{d M^2\,d t}\Lb \reg \pom \reg\, \mbox{term
; G-W contribution}\Rb +M^2 \frac{d \sigma^{\mbox{High M}}_{sd}\Lb s,M \Rb}{d M
^2\,d t}\Lb \eq{DXSSD}\Rb\,\nonumber\\
 && \,\,\,\,\,\,\,\,\,\,\,\,\,\,\,\,\,\,\,\,\,\,\,\,\,\,\,\,\,\,\,=
\,\,B_{sd} e^{- B_{sd}\,|t|} \left\{\sigma^{\mbox{High M}}_{sd}\, \frac{N^{MPSI
}_{sd}\Lb Y,Y_m; \eq{SD1}\Rb}{ N^{MPSI}_{diff}\Lb\eq{TSD}\Rb}\,+\,\sigma^{\mbox
{Low  M}}_{sd}  \right\}. \notag
\eea

 In our parametrization,
the scale of the second term is determined by the G-W mechanism.
  The fact that the $\reg \pom \reg$ term is responsible for $x_L$
behaviour, is an additional independent input.  However,  an argument
for the
 $\reg \pom \reg$ term , is that if $\Delta_\pom =0$,  then both
 $\pom\pom \pom$ and
 $\reg \pom \reg$  lead to a diffractive cross section,
 which is constant as a
function of energy.

\par

{\boldmath
\section{Predictions for LHC and Cosmic Rays Energies: $b$-dependence of  the amplitudes}
}
\fig{predi}
shows our prediction for high energy behaviour of the total, elastic and diffractive cross sections as well as
of the elastic slope.

\FIGURE[ht]{
\begin{tabular}{c c}
\epsfig{file=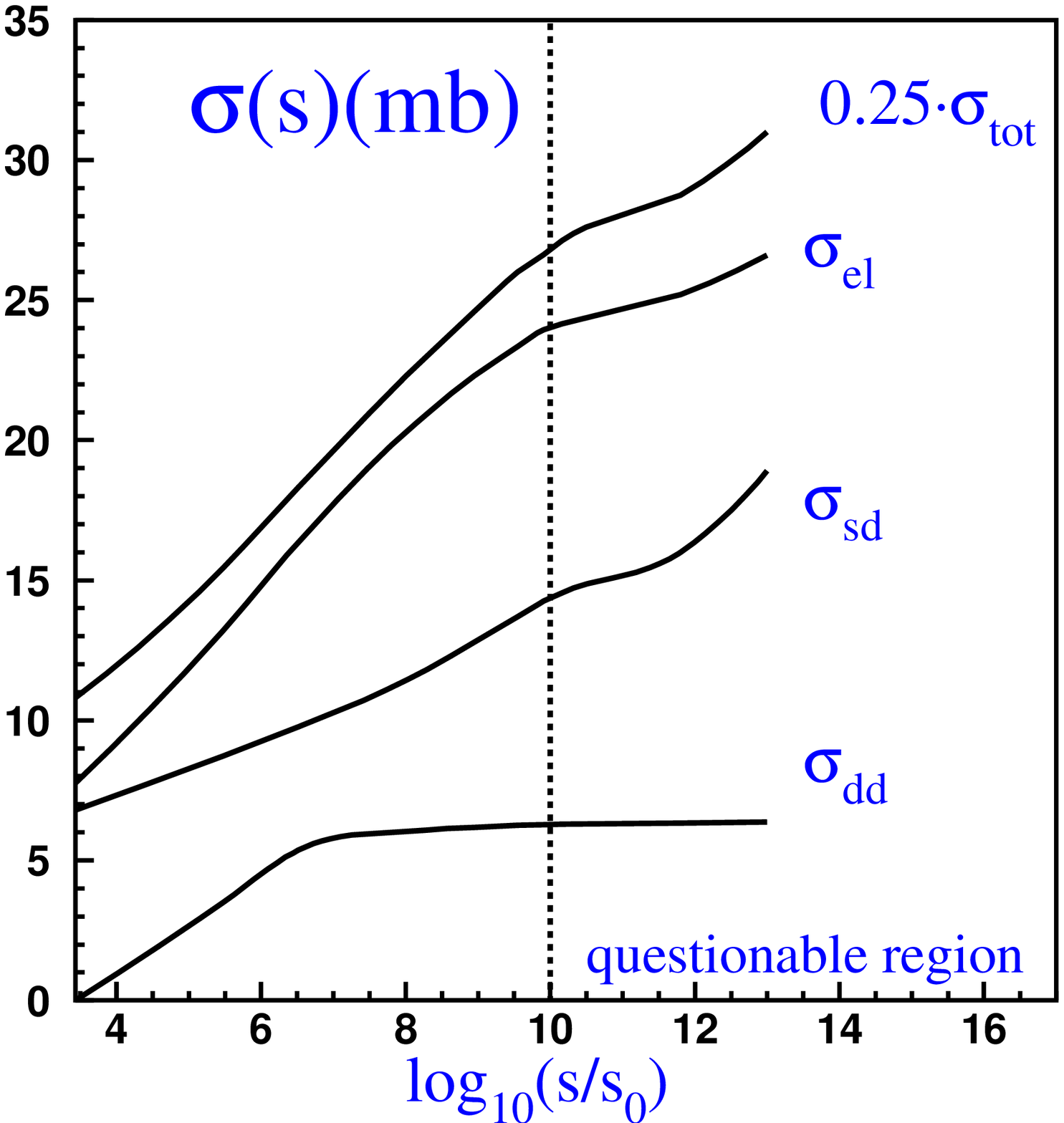,width=85mm,height=75mm} &
\epsfig{file=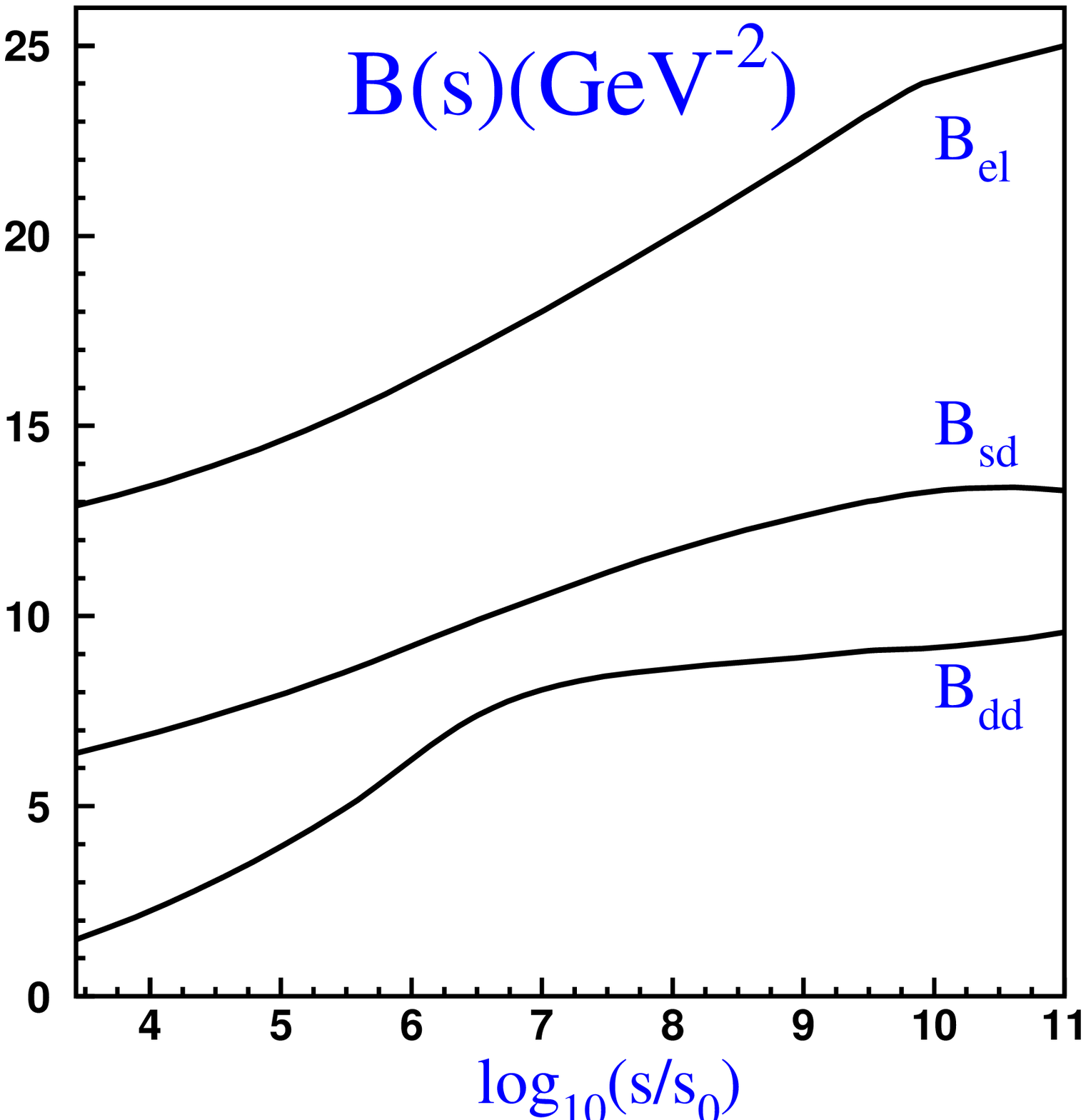,width=85mm,height=75mm}\\
\fig{predi}-a & \fig{predi}-b\\
\end{tabular}
\caption{The energy dependence of cross sections (\fig{predi} -a) and elastic
slope $B_{el}$(\fig{predi}-b)
for elastic scattering and diffractive production.
   We have  plotted $0.25\,\sigma_{tot}$, so as to show all the
predictions on the same figure.
  }
 \label{predi}
}
At W=14 $TeV$ (LHC energy): $\sigma_{tot}= 92.1\,mb$,
$\sigma_{el}=20.9\,mb$, $\sigma_{sd}=11.8 \,mb$,
$\sigma_{dd}=6.08\,mb$ and $B_{el}=20.6 \,GeV^{-2}$.
Comparing these results with  the prediction of our previous  model
 ($\sigma_{tot}=110.5\,mb$,
$\sigma_{el}=25.3\,mb$, $\sigma_{sd}=11.6 \,mb$,
$\sigma_{dd}=4.9\,mb$ and $B_{el}=20.5 \,GeV^{-2}$ at the LHC energy) we see that
the above comparison

\FIGURE[ht]{\begin{tabular}{c c}
\epsfig{file=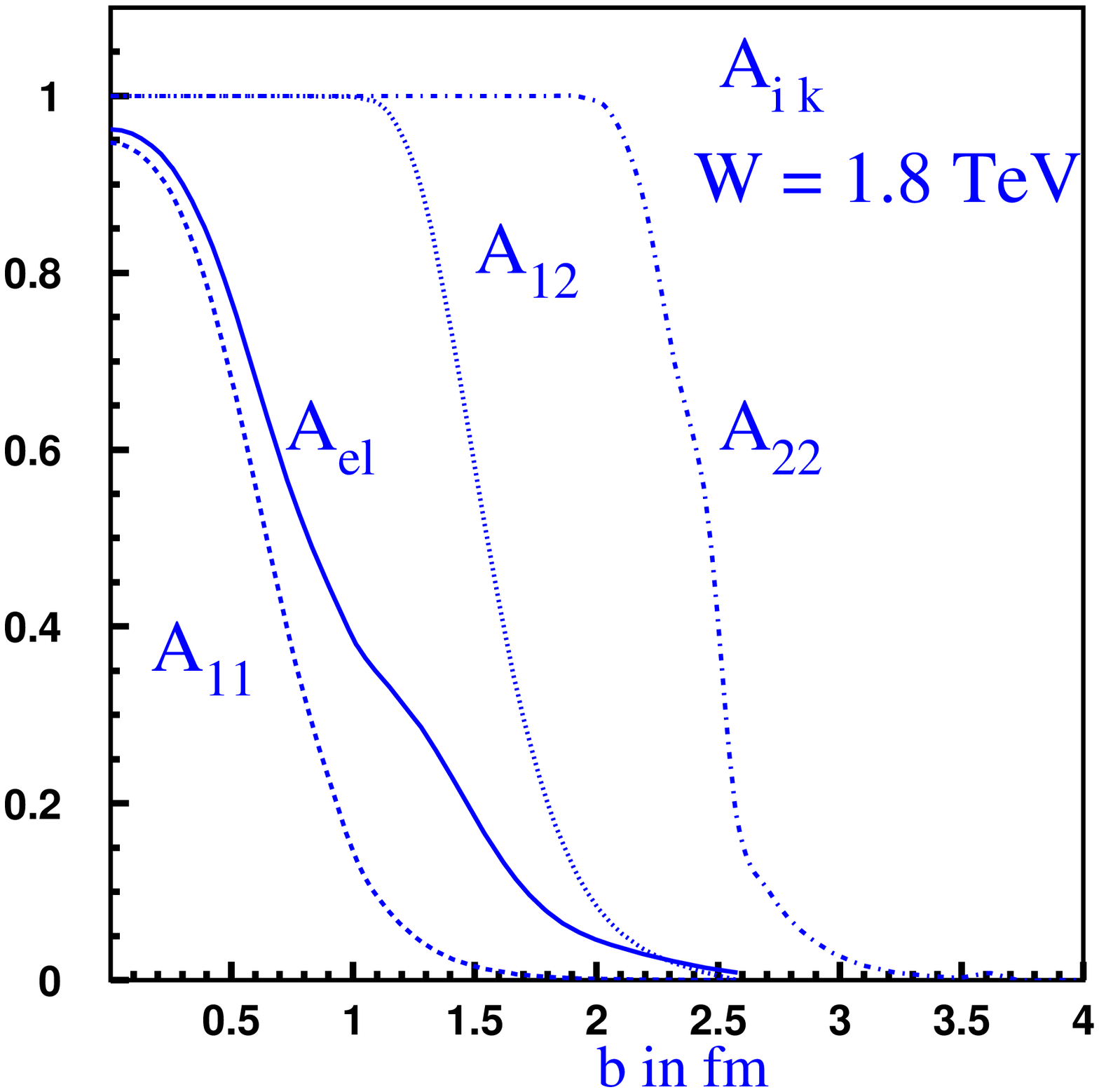,width=90mm,height=80mm}  &
\epsfig{file=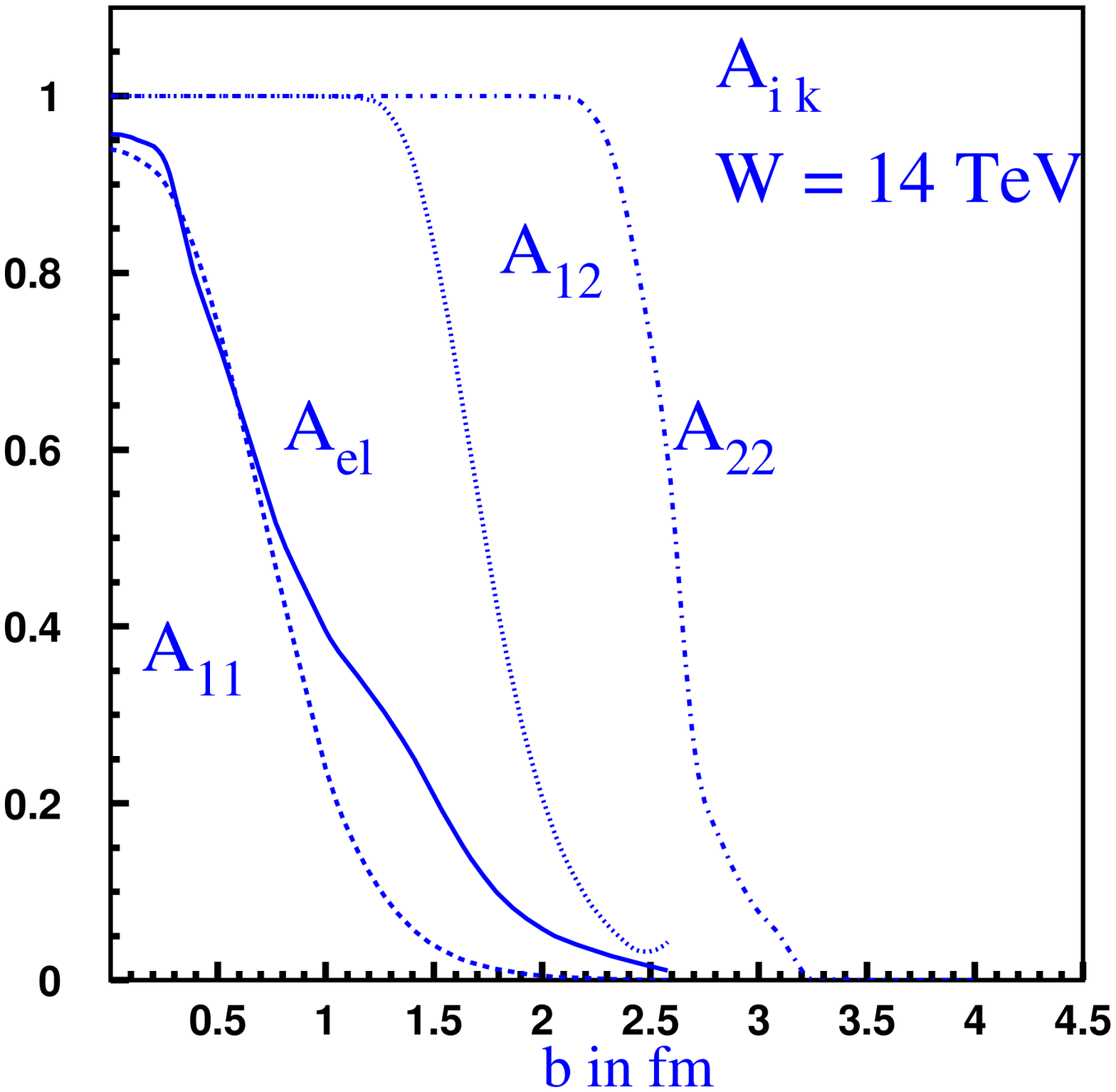,width=90mm,height=80mm} \\
\fig{am}-a & \fig{am}-b\\
\end{tabular}
\caption{Impact parameter dependence of $A_{i,k}$ and
$a_{el}$ at different energies.
\label{am}}}

leads to a few general observations, on which we
will elaborate
 in the Discussion section.\\
1) The predictions of our Ref. \cite{GLMLAST} for $\sigma_{tot}$ and
 $\sigma_{el}$ at the LHC
are considerably higher than the corresponding predictions obtained in the present analysis.
 The difference between the
two sets of predictions grow monotonically with energy.
Note, that the moderate growth of $\sigma_{tot}$ and $\sigma_{el}$
obtained in the
present study continues up to energies as high
as $10^{5}$ GeV, and probably higher. See Fig.13 and Table 3. We attribute
this behaviour to the fact, that in our fit,  $\alpha^{'}_{\pom}$ is small
but not zero.\\
2) These features reflect the fact that whereas our previous model has an
output
 compatible with
$\Delta_{eff} \approx 0.08$ up to exceedingly high energies, the present
model (in which Pomeron enhanced diagrams play a
significant role), has an output $\Delta_{eff}$ which decreases
monotonically with energy.\\
3) The qualitative features of $\sigma_{tot}$ and $\sigma_{el}$ in our two
models are
also seen in $\sigma_{sd}$. However, $\sigma_{dd}$ behaves differently,
slowly approaching a constant value above LHC energies.\\
4) Our $B_{sd}$ slope is approximately two times less $B_{el}$.
At high energy we have only the results of fit to the diffraction experimental data given in Ref. \cite{GOMO}.
The values of $B_{sd}$ from this fit are $B_{sd}= 7.7 \pm 0.6 \,( 4.2 \pm 0.5) \,GeV^{-2}$ for W = 546 (1800)\,GeV. Our values  (see \fig{predi}-b) are $B_{sd}=  8.5 (9.9)  \,GeV^{-2}$ at these energies.
The large difference in the experimental slopes have been discussed in  Ref. \cite{GOMO}.

To investigate  the quantitative features of our present model further, we
plot in Fig.14 the $ b$ dependence of
 $A_{i,k}(s,b)$ and $a_{el}(s,b)$ output amplitudes, at Tevatron and LHC
energies.
As seen, $A_{2,2}$ and $A_{1,2}$ reach the black bound at relatively low energies, whereas
$A_{1,1}$ is below the black
bound, approaching it very slowly.

As noted in Ref.\cite{GLMLAST}, the output cross sections reach the
unitarity bound when, and
only when, $A_{1,1}\; = \; A_{1,2} \; = \; A_{2,2}\; = \;1$. Since,
$A_{1,1}$ grows very slowly with energy, we
conclude that the very slow approach to the unitarity bound observed
 in Ref.\cite{GLMLAST}, also occurs in the present model.
Despite the fact that the two models have very different $t$ dependences,
 we hope that such structure of the amplitude does not depend on the model
 assumptions, but  reflects the principle features of the hadron
scattering at high energy.
 We shall expand on this issue in the Discussion
section.

\section{Survival probability of diffractive Higgs production}
In the following we  limit our discussion to the survival probability
of Higgs production, in an exclusive
central diffractive process.   Most estimates  of the values of survival
 probability have been made in the
framework of G-W mechanism, in two channel eikonal models.
A general review of such
survival probability calculations can be found in
Ref.\cite{heralhc}.
The general formulae for the calculation of the survival probability for
diffractive Higgs boson production, have been discussed
in Refs.\cite{SP2CH,heralhc,SP3P}. The structure of the survival probability
expression is shown in \fig{sp-dia}.-a.  Accordingly,
\beq \label{SP}
\langle\mid S^2_{2ch} \mid \rangle = \frac{N(s)}{D(s)},
\eeq
where,
\bea
&N(s) = \int d^2\,b_1\,d^2\,b_2
\left[\sum_{i,k} \,<p|i>^2 <p|k>^2 \,A^{i}_H(s,b_1)\,A^k_H(s,b_2) (1-A^{i,k}_S \Lb(s,(\mathbf{b}_1+\mathbf{b}_2 )
\Rb)\right]^2, \label{SP1}\\
&D(s) = \int\,d^2\,b_1\,d^2\,b_2
\left[\sum_{i,k} <p|i>^2 <p|k>^2\, A^i_H(s,b_1)\,A^k_H(s,b_2)\right]^2.
\label{SP2}
\eea
 $<p|i>$ is  equal to $\langle \Psi_{proton}\mid \Psi_i \rangle$ and , therefore, $<p|1>
= \alpha$ and $ <p|2> = \beta$.

$A_s$ denotes the soft strong interaction  amplitude given
by \eq{OMEGA},\eq{EL},\eq{SD} and \eq{DD}.
The form of $ A_H(s,b)$ has been discussed in Refs.\cite{SP2CH,heralhc}.
In our model we assume an input Gaussian $b$-dependence
for the hard  amplitudes.
\begin{equation}\label{AH}
{A_{i,k}^H} = A_H(s)\,\Gamma_{i,k}^H(b),
\end{equation}
where $A_H(s)$ is an  $s$- dependent arbitrary function which does not
depend
on $i,k$, and\\
$\Gamma_{i,k}^H(b) =
\frac{1}{\pi (R^H_{i,k})^2}\,e^{-\frac{2\,b^2}{(R^H_{i,k})^2}}$.
The hard vertices and radii ${R_{i,k}^H}^2$,
are constants derived from HERA $J/\Psi$ elastic and inelastic
photo and DIS production\cite{PSISL}.

Following Refs.\cite{SP3P,heralhc} we have introduced in the above,
two hard $b$-profiles
\bea
A^{pp}_H(b) &=&
\frac{V_{p \to p}}{2 \pi B_{el}^H} \exp \Lb-\frac{b^2}{2\,B_{el}^H} \Rb,
\label{2C10}\\
A^{pd}_H(b) &=& \frac{V_{p \to d}}{2 \pi
B_{in}^H} \exp \Lb -\frac{b^2}{2 B_{in}^H}\Rb.
\label{2C11}
\eea
 The values $B_{el}^H$=3.6 $GeV^{-2}$ and $B_{in}^H$=1 $GeV^{-2}$,
have been taken from the experimental ZEUS data on
$J/\Psi$ production at HERA (see Refs.\cite{heralhc,KOTE}).

Using \eq{EL}-\eq{DD}, the integrands of
\eq{SP1} and \eq{SP2} are reduced by eliminating common $s$-dependent
expressions,
\bea
N(s) &=& \int\,d^2 b_1 d^2 b_2
[A_H(s,b_1)\,A_H(s,b_2) (1 - A_S \Lb
\mathbf{b} = \mathbf{b}_1+\mathbf{b}_2 \Rb)]^2 \nonumber \\
&= &\int d^2 b_1 d^2 b_2\,
[(1 - a_{el}(s,b))A^{pp}_H(b_1) A^{pp}_H(b_2)
%\nonumber \\&&
 - a_{sd}(s,b)\Lb A^{pd}_H(b_1) A^{pp}_H(b_2) \right. \notag\\
& + & \left.
A^{pp}_H(b_1) A^{pd}_H(b_2) \Rb
 - a_{dd}(s,b) A^{pd}_H(b_1) A^{pd}_H(b_2)]^2
\label{SP3},
\eea
\beq \label{SP4}
D = \int d^2 b_1 d^2 b_2
\left[A^{pp}_H(b_1) A^{pp}_H(b_2)\right]^2.
\eeq
\par

%%%%%%%%%%%%%%%%%%%`Survival probability in diffractive Higgs production in high density QCD,%%%%%%%%%%%%%%%%%%%%%%%%%%%%%%%%%%%%%%%%%%%%%%%%%%%%%
\FIGURE[ht]{
\centerline{\epsfig{file=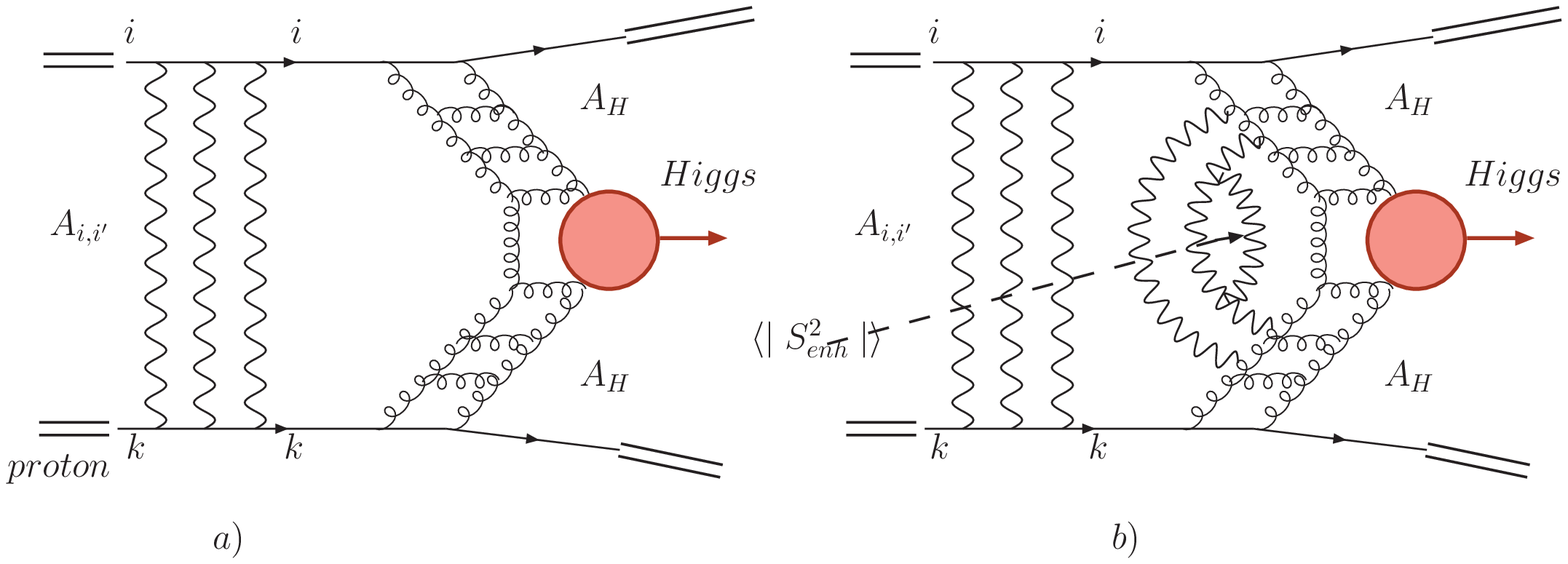,width=150mm}}
\caption{Survival probability for exclusive central diffractive
production of the Higgs boson. \fig{sp-dia}-a shows the contribution to
the survival probability in
the G-W mechanism, while \fig{sp-dia}-b  illustrates the origin to the
additional factor
$\langle\mid S^2_{enh}\mid\rangle$. }
\label{sp-dia}}
%%%%%%%%%%%%%%%%%%%%%%%%%%%%%%%%%%%%%%%%%%%%%%%%%%%%%%%%%%%%%%%%%%%%
\par
\eq{SP} does not give a correct estimate for the survival
probability,
and should be multiplied by a factor ($\langle\mid S^2_{enh} \mid \rangle$)  that,
 incorporates the possibility for the Higgs boson to be emitted from
the enhanced diagrams (see \fig{sp-dia}-b). Therefore, the resulting survival
 probability can be written as
\beq \label{SPF}
\SP \,\,\,=\,\, \langle\mid S^2_{enh} \mid \rangle\Lb \mbox{\eq{SPE}}\Rb\,
 \times\,\langle\mid S^2_{2ch} \mid \rangle\Lb \mbox{\eq{SP}}\Rb.
\eeq

The first attempt to find  $ \langle\mid S^2_{enh} \mid \rangle$
was made
in Refs.\cite{JM,KKLM},
where this factor was calculated neglecting the fact that the
   Higgs boson
could be
 produced from the two gluon scattering with a  difference in rapidity
$\delta Y_H = \ln (M^2_H/s_0)$ (see for example Refs. \cite{DUR,JM}).
The MPSI approach for this case is shown in \fig{spen1}, and it leads to
the  result:
\beq \label{SPE1}
\langle\mid S^2_{enh} \mid \rangle\Lb Y\Rb\,\,=\,\,\frac{\partial}{\partial \,T} N^{MPSI}_{el}\Lb T \Rb\,=\,
\frac{1}{T^3(Y)}\left\{ - T(Y) \,+\,e^{\frac{1}{T(Y)}}\,\Lb 1  +  T(Y) \Rb \,\Gamma_0\Lb \frac{1}{T(Y)}\Rb\right\},
\eeq
where $ N^{MPSI}_{el}\Lb T \Rb$  is given \eq{AMMPSI}. It was
originally suggested
 to divide  this factor by
 $\langle\mid  S^2_{enh}\mid \rangle\Lb Y = \delta Y_H \Rb $,

\beq \label{SPE2}
\langle\mid S^2_{enh} \mid \rangle\Lb Y\Rb\,\,=\,\,\frac{\langle\mid S^2_{enh}
\mid \rangle\Lb Y; \eq{SPE1}\Rb}{\langle\mid S^2_{enh} \mid \rangle\Lb Y=\delta
 Y_H; \eq{SPE1}\Rb}.
\eeq
In this paper we take into account $\delta Y_H$ in a consistent way,
 which was outlined in Ref. \cite{KKLM}, and presented in \fig{spen2}
[16]. At
  rapidity $ Y - Y'  - \h \delta Y_H$, one of the partons (Pomerons) will
produce a Higgs
boson, and it
 should be removed from the cascade evolution. Therefore, those  partons which
will participate in the evolution will
  be characterized by a new generating function,
\bea\label{SPE3}
&& \tilde{Z}\Lb Y - Y' - \h \delta Y_H,u \Rb\,\,\,= \\
&& e^{ - \Delta_\pom ( Y - y'- \h \delta Y_H)}\,\frac{\partial
Z^{MFA}\Lb Y - Y' - \h \delta Y_H,u\Rb}{\partial\,u}\,\,=\,\,\frac{1}{\Lb u \,+\,(1 - u) \,e^{
- \Delta_\pom ( Y - Y'- \h \delta Y_H)} \Rb^2}. \notag
\eea
$\tilde{Z}\Lb Y - Y'- \h \delta Y_H; u \Rb$ should be evolved to rapidity $ Y -
 Y' $ using \eq{FDEQ}. This evolution results in an improved  generating
function
\beq \label{SPE4}
\widetilde{\widetilde{Z}}\Lb Y - Y' ,u \Rb\,\,=\,\,\frac{\Lb u + (1 - u)\,e^{\h \Delta_\pom \delta Y_H} \Rb^2}{
 \Lb u + (1 - u)\,e^{\Delta_\pom ( Y - Y) }\Rb^2}.
\eeq
 Note that for central Higgs boson production $\h Y = Y' $.

Symmetrically, using \eq{SPE3}
 we need to find $ \tilde{Z}\Lb Y'- \h \delta Y_H ,u \Rb$, and from
\eq{SPE4},
 $\tilde{Z}\Lb Y' ,u \Rb$. The result for this function
 is
\beq \label{SPE41}
\widetilde{\widetilde{Z}}\Lb Y',u \Rb\,\,\,=\,\,\,\frac{\Lb u + (1 - u)\,e^{\h \Delta_\pom \delta Y_H} \Rb^2}{
 \Lb u + (1 - u)\,e^{\Delta_\pom Y'  }\Rb^2}.
\eeq

Using these  generating functions we obtain
\bea
&&\langle \mid S^2_{enh}\Lb MPSI\Rb \mid \rangle \left(Y\right)\,\,\,=  \notag\\
&& \,\,\,\,\,\,\,\,\,\,\,\,\,\,\,\,\,\,\,\,\,\,\,\,=\sum^{\infty}_{n=1}\,\,\frac{(-1)^{n + 1}}{n!}\,
\,\,\gamma^n\,\,  \frac{\partial^n\,\widetilde{\widetilde{Z}}(Y -Y' ,u^p)}{\partial^n\, u^p}|_{u^p = 1}
\,\, \frac{\partial^n\,\widetilde{\widetilde{Z}}(Y',u)\,u^t)}{\partial u_t}|_{u^t = 1}\label{SPE}\\
 &&\,\,\,\,\,\,\,\,\,\,\,\,\,\,\,\,\,\,\,\,\,\,\,\,= \,\,S\Lb {\cal T}( Y)\Rb \,\,-\,\,2\,e^{ - \Delta_\pom (Y - \delta Y_H)/2}\,S1\Lb {\cal T}( Y)\Rb\,\,+\,\,
e^{ - 2\Delta_\pom (Y - \delta Y_H)/2}\,S2\Lb {\cal T}( Y\Rb; \notag\\
&& \,\,\,\,\,\,\,\,\,\,S(T) = \frac{1}{T^3}\left\{ - T\,+\,e^{\frac{1}{T}}\,\Lb 1  + T \Rb \,e^{\frac{1}{T}}\,\Gamma_0\Lb \frac{1}{T}\Rb\right\}; \label{SPE51}\\
&& \,\,\,\,\,\,\,\,\,S1(T) = \frac{1}{T^3}\left\{   -T (1 + T) \,+\, ( 1 + 2T)\, e^{\frac{1}{T}}\,\Gamma_0\Lb \frac{1}{T}\Rb\right\};\label{SPE52}\\
&& \,\,\,\,\,\,\,\,\,S2(T) = \frac{1}{T^3} \left\{ T\left[ \Lb  T  - 1 \Rb^2 - 2 \right]  \,+\, ( 1 + 3 T)\, e^{\frac{1}{T}}\,\Gamma_0\Lb
\frac{1}{T}\Rb\right\}.\label{SPE53}
\eea
where
\beq \label{TT}
{\cal T} \Lb Y\Rb\,\,\,=\,\,\gamma \Lb e^{  \Delta_\pom ( Y - Y')} \,-\, 1\Rb\,
\Lb e^{ \Delta_\pom Y'} \,-\, 1\Rb.
\eeq

\DOUBLEFIGURE[ht]{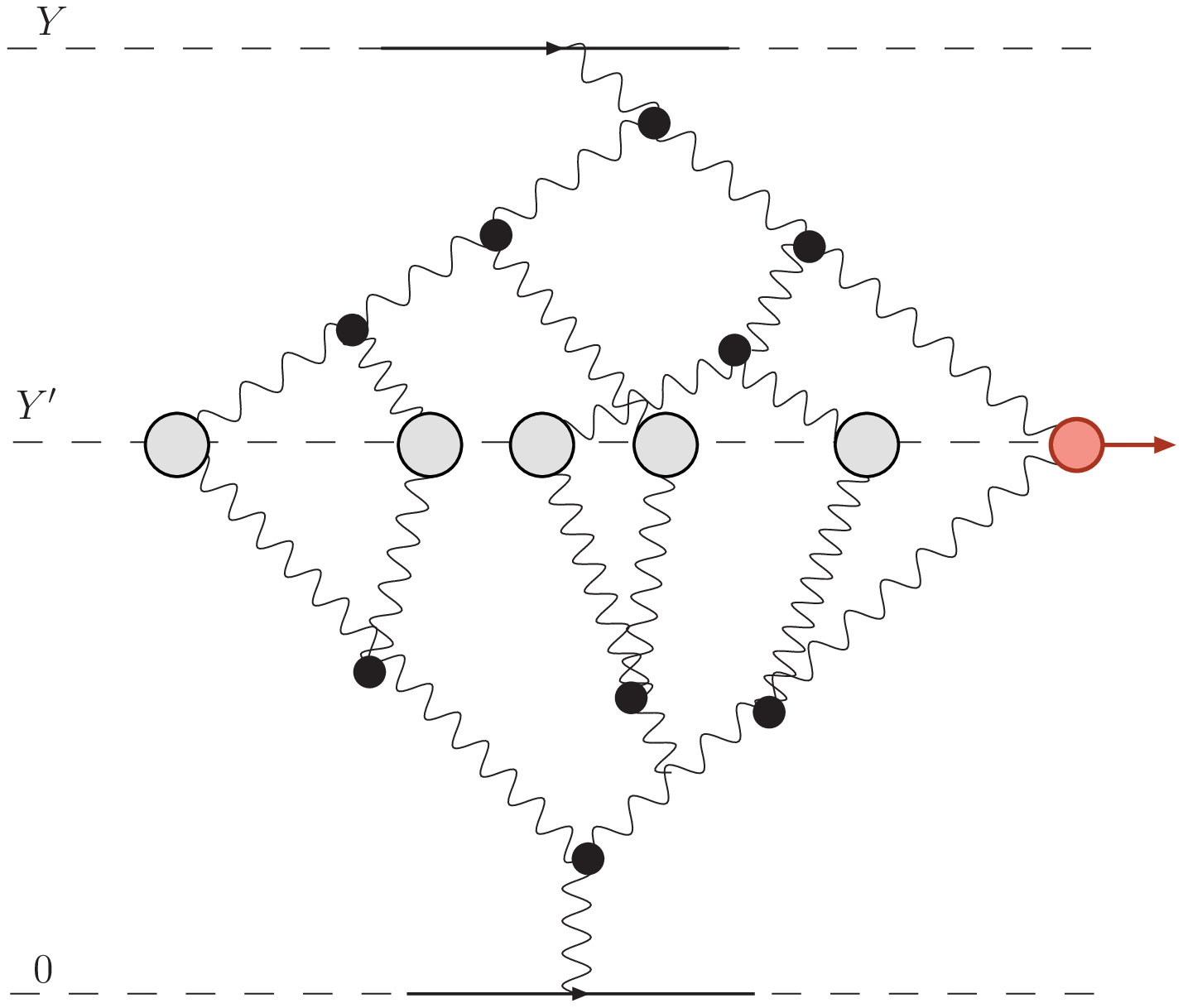,width=85mm,height=68mm}{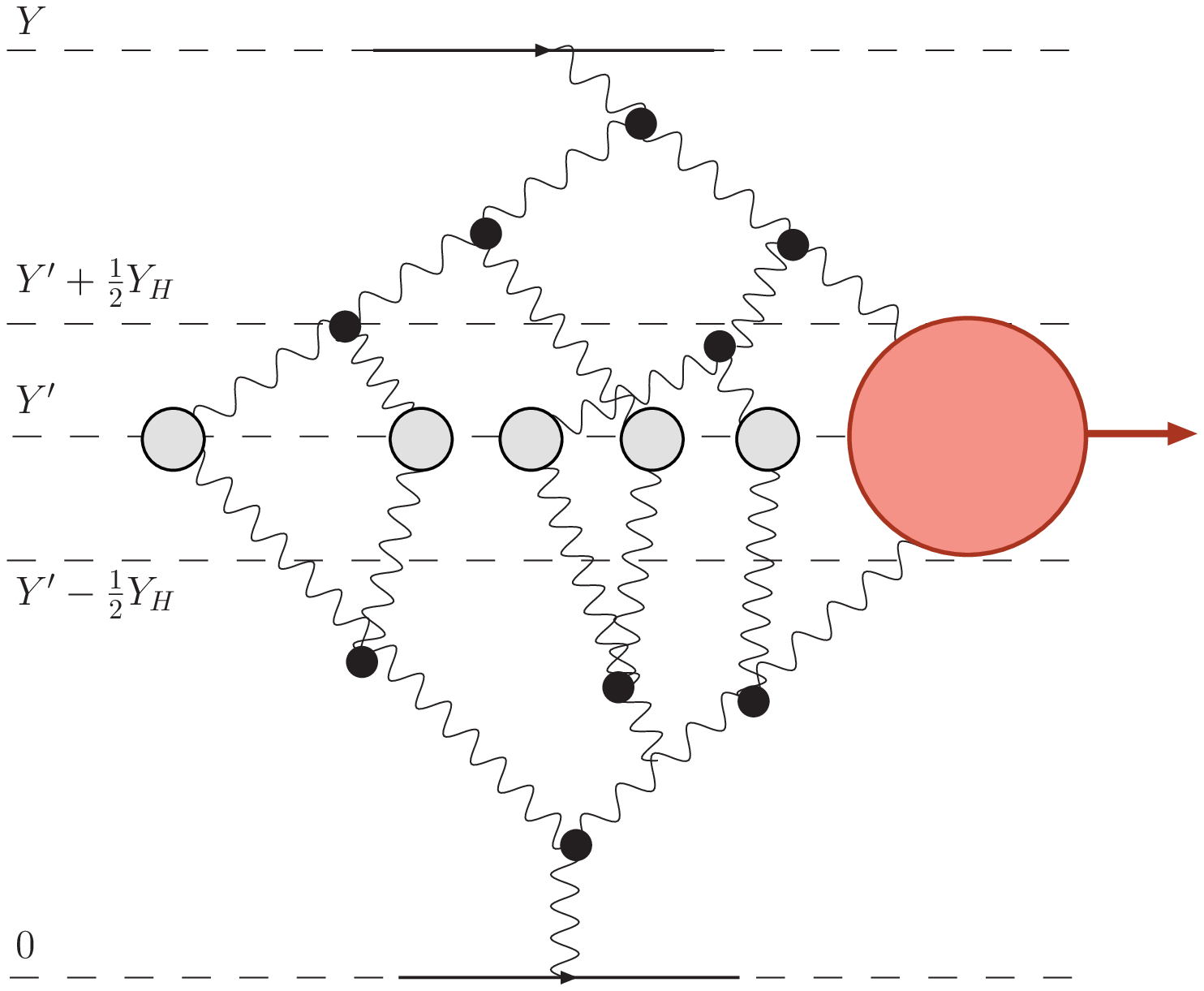,width=85mm,height=68mm}
{MPSI approach for $\langle \mid S^2_{enh}\mid \rangle$ in the case when we neglect $\delta Y_H = \ln(M^2_H/s_0)$.
\label{spen1}}
{MPSI approach for $\langle \mid S^2_{enh}\mid \rangle$ in the case when we  take into account $\delta Y_H = \ln(M^2_H/s_0)$.
\label{spen2}}

%%%%%%%%%%%%%%%%%%%%%%%%%%%%%%%%%%%%%%%%%%%%%%%%%%%%%%%%%%%%%%%%%%%%%%%%
\FIGURE[h,t]{
\centerline{\epsfig{file= 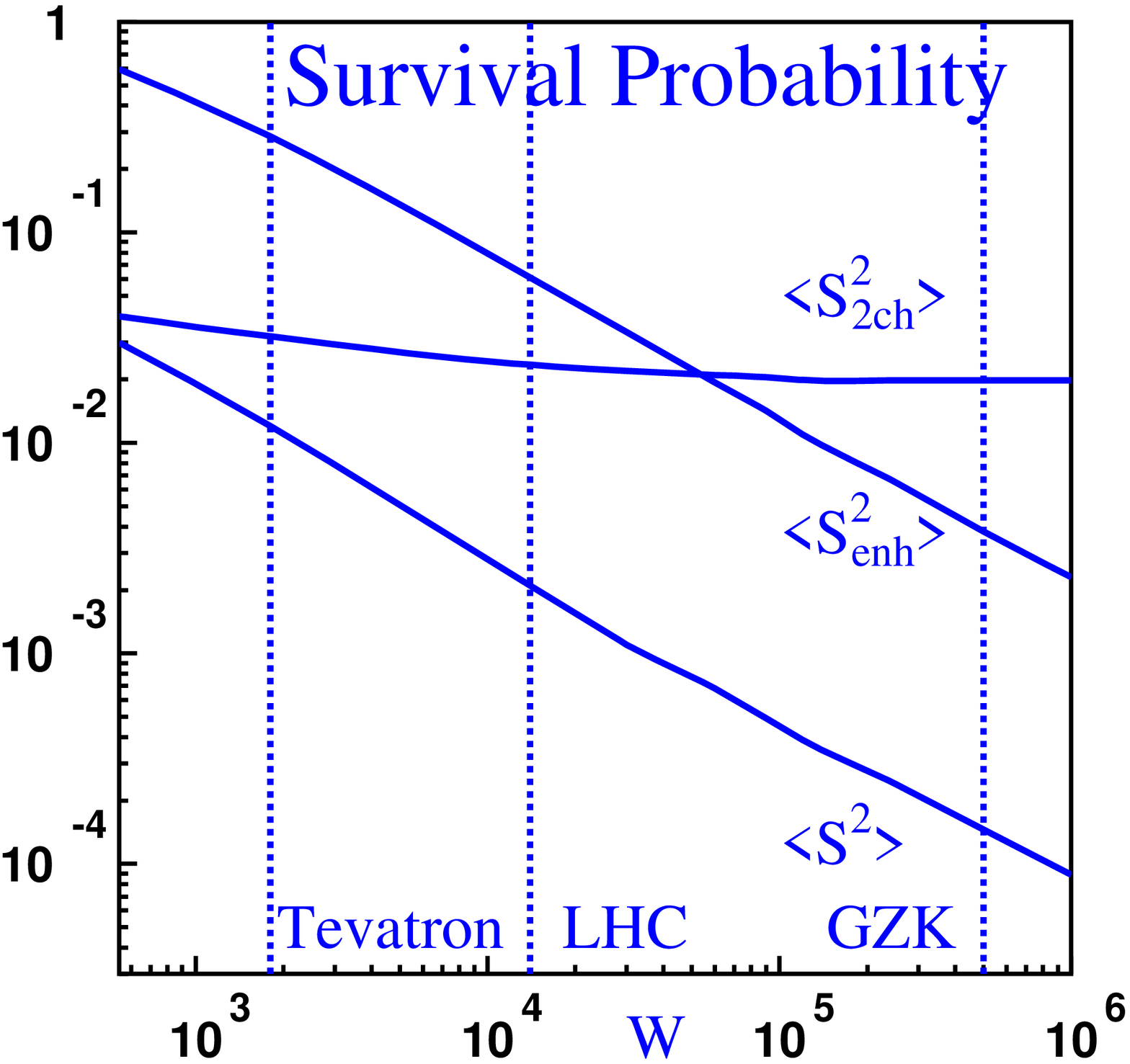,width=160mm,height=130mm}}
\caption{Energy dependence of centrally produced Higgs survival probability.}
\label{sp}}

Using \eq{SPF}-\eq{SPE53}, we calculate the survival probability  $\langle
\mid S^2\mid \rangle$ for
exclusive Higgs production in central diffraction. Our results
are plotted in \fig{sp}.
 In the following we focus on a detailed discussion of our LHC predictions
based on the
above. Calculating in the framework of a two amplitude eikonal model,
$\langle \mid S^2_{2ch}\mid \rangle$  for exclusive Higgs production
 in central diffraction
is equal to 2.35\%,  which is close to the value of this part of
the survival
 probability estimated by the Durham group \cite{KMROLD,KMRNEW}, and by
our
group \cite{heralhc}.
However, the additional factor $\langle \mid S^2_{enh}\mid \rangle$ =
0.063
 as derived from  \eq{SPE},
  leads to a final $\SP = 0.15 \%$. This result
  reflects a tendency for the
value of the survival probability to be much smaller, than when evaluated
in
the two channel models \cite{JM,GLMLAST}.
 Note,  that
 the approximate expression for $\langle \mid S^2_{enh}\mid \rangle$
derived from  \eq{SPE1}
with our parameters,  turns out
to be   three times smaller. In Ref. \cite{KMRNEW} the Pomeron diagrams were taken into account, but
nevertheless, their contribution in the calculation of $\SP$ was
 omitted, without any explanation.

\par

%%%%%%%%%%%%%%%%%%%%%%%%%%%%%%%%%%%%%%%%%%%%%%%%%%%%%%%%%%%%%%%%%%%%%%%%%%%%%%%%

\TABLE[ht]{
\begin{tabular}{||l|l|l|l||}
\hline \hline
  &  \,\,\,\,\,\,\,\,\,\,\,\,\,\,\,\,\,\,\,\,\,\,\,\,Tevatron & \,\,\,\,\,\,\,\,\,\,\,\,\,\,\,\,\,\,\,\,\,\,\,\, LHC &\,\,\,\,\,\,\,\,\,\,\,\,\,\,\,\,\,\, $W= 10^5 GeV$  \\
 & GLM\,\,\,\,\,\,\,\,\,\,\,\,\,\,\,\,\,\,\,\,\,\,\,\,KMR & GLM\,\,\,\,\,\,\,\,\,\,\,\,\,\,\,\,\,\,\,\,\,\,\,\,KMR & GLM\,\,\,\,\,\,\,\,\,\,\,\,\,\,\,\,\,\,\,\,\,\,\,\,KMR\\\hline
$\sigma_{tot}$( mb ) & 73.29 \,\,\,\,\,\,\,\,\,\,\,\,\,\,\,\,\,\,\,\,\,\,\,\,74.0 & 92.1\,\,\,\,\,\,\,\,\,\,\,\,\,\,\,\,\,\,\,\,\,\,\,\,\,\,88.0 & 108.0\,\,\,\,\,\,\,\,\,\,\,\,\,\,\,\,\,\,\,\,\,\,\,\,98.0 \\ \hline
$\sigma_{el}$(mb) & 16.3 \,\,\,\,\,\,\,\,\,\,\,\,\,\,\,\,\,\,\,\,\,\,\,\,\,\,\,16.3 & 20.9 \,\,\,\,\,\,\,\,\,\,\,\,\,\,\,\,\,\,\,\,\,\,\,\,20.1 & 24.\,\,\,\,\,\,\,\,\,\,\,\, \,\,\,\,\,\,\,\,\,\,\,\,\,\,\,22.9 \\\hline
$\sigma_{sd}$(mb) & 9.76 \,\,\,\,\,\,\,\,\,\,\,\,\,\,\,\,\,\,\,\,\,\,\,\,\,\,10.9 & 11.8 \,\,\,\,\,\,\,\,\,\,\,\,\,\,\,\,\,\,\,\,\,\,\,\,13.3 &14.4  \,\,\,\,\,\,\,\,\,\,\,\,\,\,\,\,\,\,\,\,\,\,\,\,15.7 \\
\,\,\,\,\,\,\,\,\,\,\,\,\,\,\,\,\,\,\,\,\,\,\,\,\,\,\,\,\,\,\,\,$\sigma^{\mbox{low M}}_{sd}$ & 8.56\,\,\,\, \,\,\,\,\,\,\,\,\,\,\,\,\,\,\,\,\,\,\,\,\,\,\,4.4 & 10.52\,\,\,\, \,\,\,\,\,\,\,\,\,\,\,\,\,\,\,\,\,\,
5.1 & 12.2\,\,\,\, \,\,\,\,\,\,\,\,\,\,\,\,\,\,\,\,\,\,\,\,\,\,\,\,5.7 \\
\,\,\,\,\,\,\,\,\,\,\,\,\,\,\,\,\,\,\,\,\,\,\,\,\,\,\,\,\,\,\,\,$\sigma^{\mbox{high  M}}_{sd}$ & 1.2\,\,\,\, \,\,\,\,\,\,\,\,\,\,\,\,\,\,\,\,\,\,\,\,\,\,\,\,\,\,\,6.5 & 1.28\,\,\,\, \,\,\,\,\,\,\,\,\,\,\,\,\,\,\,\,\,\,\,\,\,\,\,8.2 & 2.2\,\,\,\,\,\,\,\,\,\,\,\,\,\,\,\, \,\,\,\,\,\,\,\,\,\,\,\,10.0 \\ \hline
$\sigma_{dd}$(mb) & 5.36
\,\,\,\,\,\,\,\,\,\,\,\,\,\,\,\,\,\,\,\,\,\,\,\,\,\,7.2 &  6.08 \,\,\,\,\,\,\,\,\,\,\,\,\,\,\,\,\,\,\,\,\,\,\,\,13.4 &
6.29 \,\,\,\,\,\,\,\,\,\,\,\,\,\,\,\,\,\,\,\,\,\,\,\,17.3 \\
 \hline
$\Lb\sigma_{el} + \sigma_{sd} + \sigma_{dd}\Rb/\sigma_{tot}$ & 0.428 \,\,\,\,\,\,\,\,\,\,\,\,\,\,\,\,\,\,\,\,\,\,\,\,0.464 & 0.421 \,\,\,\,\,\,\,\,\,\,\,\,\,\,\,\,\,\,\,\,\,\,\,\,0.531 & 0.412\,\,\,\,\,\,\,\,\,\,\,\,\,\,\,\,\,\,\,\,\,\,\,\,\,\,0.57 \\ \hline
$S^2_{2ch}(\%)$ & 3.2  \,\,\,\,\,\,\,\,\,\,\,\, \,\,\,\,\,\,\,\,\,\,\,\,\,\,2.7 - 4.8 & 2.35  \,\,\,\,\,\,\,\,\,\,\,\, \,\,\,\,\,\,\,\,\,\,1.2-3.2 &
2.0  \,\,\,\,\,\,\,\,\,\,\,\, \,\,\,\,\,\,\,\,\,\,\,\,\,\, 0.9 - 2.5 \\
\hline
$S^2_{enh}(\%)$ & 28.5 \,\,\,\,\,\,\,\,\,\,\,\, \,\,\,\,\,\,\,\,\,\,\,\,100 & 6.3  \,\,\,\,\,\,\,\,\,\,\,\, \,\,\,\,\,\,\,\,\,\,\,100 & 3.3 \,\,\,\,\,\,\,\,\,\,\,\, \,\,\,\,\,\,\,\,\,\,\,\,\,\,100 \\ \hline
$S^2(\%)$ & 1.2  \,\,\,\,\,\,\,\,\,\,\,\, \,\,\,\,\,\,\,\,\,\,\,\,\,\,2.7 - 4.8 & 0.21 \,\,\,\,\,\,\,\,\,\,\,\, \,\,\,\,\,\,\,\,\,\,1.2-3.2 &
0.05\, \,\,\,\,\,\,\,\,\,\,\,\, \,\,\,\,\,\,\,\,\,\,\,\,0.9 - 2.5\\
\hline \hline
\end{tabular}
  \caption{Comparison of  the GLM ( this paper) and KMR\cite{KMRNEW}
models.  }
\label{t3}}

{\bf
\section{Discussion}
}

 In this paper we have presented an approach  for soft interactions at high
energies based on two ingredients:\\
1) The Good-Walker mechanism for elastic and low mass diffraction.\\
2) Pomeron enchanced contributions which leads to the exact Pomeron
Green's
function, which is significantly different from one Pomeron exchange. This component provides the main contribution
to high mass diffraction.

\par
Our enhanced Pomeron formalism, is based on the observation that the soft
scattering cross sections and slopes, can be reproduced with
$\alpha^{'}_{\pom} \approx  \; 0.01\; GeV^{-2}$,
rather than $\alpha^{'}_{P} =  \; 0.25\; GeV^{-2}$ typical of
conventional Regge phenomenology. Our result of a very small
$\alpha^{'}_{\pom}$, implies that the observed shrinkage of the forward
differential cross sections, traditionally associated with
$\alpha^{'}_{\pom} \; > \;0$, can be also  reproduced  with
$\alpha^{'}_{\pom} \; \approx \;0$ coupled to strong screening, which
produces the desired shrinkage.

\par
In the following we  discuss further the properties of our model, by
comparing it with the KMR model \cite{KMRNEW}, which is conceptually similar to
ours, having the same two mechanisms: G-W  and multi-Pomeron interactions.
%\subsection{Comparison with the KMR model}
 In the KMR model $\alpha'_{\pom}\; \equiv \; 0$ is an input
assumption. In
the present GLM model  $\alpha'_{\pom}$ is a fitted parameter, and its
value  $\alpha'_{\pom}\; = \; 0.012 $ is an output.
  As we have discussed the small value of $\alpha'_{\pom}$, led us to
our key hypothesis that we advocate in this paper:
  the soft processes are not so soft, but stem from  short
distances, where
the QCD coupling is small ($\as \approx 0.12$ to $ 0.16$).  Using this
hypothesis we built
our  theoretical  approach. This approach is  self consistent
based on
pQCD. This enables us to restrict our summation only to triple Pomeron
vertices. KMR summation is based on an {\it ad hoc} assumption
\beq \label{KMR1}
\Gamma\Lb n\,\,\, Pomerons \to m\,\,\, Pomerons\Rb \,\,\propto  \,\,
n\,m\,\lambda^{ m + n -2}.
\eeq
Further, KMR claim that  \eq{KMR1} leads to their main equation
(Eq.(26) of Ref.\cite{KMRNEW}), which corresponds to the parton
model. Both claims
maybe correct, but they have not been proven in the KMR paper. A formal
difficulty also noted by the KMR authors (at the beginning of their
section 4.3) is that the formulae for diffractive production are {\it ad
hoc},  and
are not actually compatible with  our \eq{KMR1} and Eq.(26) of
Ref.\cite{KMRNEW}.

In spite of the fact, that we do not think that KMR model is able to
provide
 reliable
 estimates, it is interesting to compare our results, since both are based
on the
same physics.\\
1) The introduction of  Pomeron induced interactions
 to the calculation,
 results in the
accumulation of Pomeron loops along the initial Pomeron propogator, which
lead to a monotonic reduction of the output $\Delta_{eff}$ with energy.
In the KMR model, where additional diagrams, and
not only  diagrams with triple Pomeron interactions
have been included, this process occurs more rapidly  than in our
model. Accordingly, they
choose a very high input value,  $\Delta_{in} \; = \; 0.55$. In our fit
we have  $\Delta_{in} \; = \; 0.335$.  Both in the KMR model and in our
 model the
 effective shrinkage of the diffraction peak, stems from the Pomeron
interactions.
The difference is that  $R^{2}(s)$ in the GLM model  grows as
$ln^{2}s$ with a
coefficient proportional to  $\alpha'_{P}$.
Since in evaluating our summations, we have made an approximation in which
$\alpha'_{\pom}$ = 0, our calculations are only trustworthy up to
  $W  \approx 10^5 \,GeV$.
 The high energy output of GLM and KMR models are
presented in Table 3.
For $\sigma_{dd}$ the contribution to the
 diffractive channels coming
from the Pomeron enhanced diagrams, are larger in KMR than in GLM.
 Note that $\sigma_{dd}$,  as calculated by GLM,
saturates just above LHC energies, while in KMR it continues growing even
at energies of W = $10^{5}$ GeV, where it is predicted to be much larger
than the $\sigma_{sd}$.\\
2) In Table 3 we define $\sigma^{\mbox{Low M}}_{sd}$ as the contribution of G-W mechanism, while in the KMR model low and high mass  diffraction  is allocated to mass values $M < M_0 = 2.5 \,GeV$ and $M > M_0 = 2.5 \,GeV$.\\
3) The behaviour of the ratio
$R_{D} \; = \; \frac{(\sigma_{el} + \sigma_{sd} +
\sigma_{dd})}{\sigma_{tot}}$ conveys information regarding the onset of
unitarity constraints at high enough energy. In the G-W model the Pumplin
bound $R_{D}\;\leq\; 0.5$ is relevant. The multi-Pomeron induced
contributions are not included in this bound. In the GLM model,
 $R_{D}\;\ < \; 0.5$ and decreases very slowly with energy. It's
corresponding $R^{G-W}_{D} \approx \; 0.35$, is a constant similar to the
output of Ref.\cite{GLMLAST}. In KMR,  $R_{D}\;\ > \; 0.5$ and grows with
energy.
 The difference between GLM and KMR appears to be due to the different
summations of the Pomeron induced diagrams, incorporated in the models.\\
4) The difference in the values of the survival probabilities calculated
in the two channel models are significant. The KMR estimate should be
reduced by $S^{2}_{enh}$ which is ignored in their calculation. Including
this factor  brings the GLM and KMR numbers closer.\\
5) The data analysis aimed at determining the three opacities
$\Omega_{i,k}$, is conceptually different in the GLM and KMR models. In
principle this information should be obtained utilizing
Eq.(2.12)-Eq.(2.18).  However, the experimental information we have at the
UA(4)-Tevatron energies is not sufficient to constrain the Pomeron
parameters. In the GLM model we, thus, included both the Pomeron and
Reggeon trajectories, covering also the extensive ISR data. This enabled
us to extract the Pomeron (fitted) parmeters.
KMR have adopted a different strategy, obtaining the Pomeron opacities
from a  good fit to  $\frac{d\sigma_{el}}{dt}$ in UA(4)-Tevatron energy range. This implies a good
reproduction of $\sigma_{tot}$, $\sigma_{el}$ and $B_{el}$.
From their
paper \cite{KMRNEW}, it is not clear whether  KMR also fit the diffractive
channels. This is not the only ambiguity in the  KMR
presentation of their results. Some of their parameters are explicitly
presented and fitted, some are presented and "tuned", some are assumed
and some are implied, but not explicitly  presented.\\
6)
As noted, we attribute our dynamical result to the output of our fit, in
which $\alpha'_{\pom}$ is small but not zero! Both our model and KMR,
predict total and elastic cross sections which are significantly smaller
than in two channel eikonal estimates, which do not include the Pomeron enhanced contributions (see
Refs. \cite{2CH,KMROLD,GLMLAST}).
 The differences are sufficiently large,
  so that measurements at LHC and Auger should be able to discrimenate
between the
various approaches. See details below.\\
7)
The most practical, and perhaps the most interesting result we have
obtained, is the small value (about 0.15\%) for the survival probability
of central diffractive Higgs production at the LHC. The very small value
of the final $S^{2}_{H}$, is due to the smallnes of $S^{2}_{2ch}$,
multiplied by the small $S^{2}_{enh}$. Our calculation does not include
further reductions of $S^{2}_{H}$, due to additional short distance
processes (see Refs.\cite{BBK,JM}). These may further reduce the result we
obtained for the survival probability.

 The region of applicability for our formulae is given by \eq{KR},
 in which,  $Y \ll 1/\gamma = 41$,  and by the fact that in our procedure
for summing Pomeron diagrams
 we considered $\alpha'_{
\pom} = 0$.  The first restriction leads to a very large kinematic
range for rapidity. The seco
that
 $\alpha'_\pom \,\ln(s/s_0) \,<\,0.25/m^2_1$
 in our parameterization,
 leads to $Y < 25$ . This  region, is  marked as  questionable
in
\fig{predi}-a.   The set of formulae
   in section 3.6 do not include the contribution of
the 'fan'
 diagram to the elastic amplitude. In our context, it is important that
the
contribution of such diagrams which  determines the high mass diffraction,
turns  out to be small.
%\subsection{Experimental signatures}
 In the following we list a few experimental signatures, which should be
measured at the LHC in the near future. These will give us a clue
regarding the veracity of the  models discussed in the paper:\\
i)  Measurements of $\sigma_{tot}$ and $\sigma_{el}$ should serve as a
critical test for the relevance of Pomeron enhanced diagrams. The
difference in predictions for models including (excluding) Pomeron
enhanced diagrams, becomes even more significant at Cosmic Ray
energies. The Auger experiment, where we expect results for cross sections
in the near future, at energies in the $10^{5}$ GeV range, should allow us
to discriminate between the alternative approaches.\\
ii) The Pomeron enhanced contribution to the diffractive channels, as
calculated by KMR, is considerably larger than our predictions. This is
significant for $\sigma_{dd}$, which acquires a large value in the KMR
approach.\\
iii) An early estimate of the value of $S^{2}_{H}$ should be obtained by
an
LHC measuremnt of the role of hard central LRG dijet production in a
GJJG configuration, when compared to the pQCD prediction.

To summarize,   we  developed an approach which  is self consistent,
and is
 based on a perturbative QCD input. We hope that a more microscopic
approach
with roots in QCD saturation, can be built and
we believe that this paper will contribute to such  an effort.

%%%%%%%%%%%%%%%%%%%%%%%%%%%%%%%%%%%%%%%%%%%%%%%%%%%%%%%%%%%%%
\section* {Acknowledgements}
%%%%%%%%%%%%%%%%%%%%%%%%%%%%%%%%%%%%%%%%%%%%%%%%%%%%%%%%%%%%%%
We are grateful to Omry Netzer and Andrey Kormilitzin for useful  discussions on the subject.
We thank Michail Ryskin for  an interesting correspondance and  criticism.
 This research was supported
in part by the Israel Science Foundation, founded by the Israeli Academy of Science
and Humanities, by BSF grant $\#$ 20004019 and by
a grant from Israel Ministry of Science, Culture and Sport and
the Foundation for Basic Research of the Russian Federation.

%%%%%%%%%%%%%%%%%%%%%%%%%%%%%%%%%%%%%%%%%%%%%%%%%%%

%%%%%%%%%%%%%%%%%%%%%%%%%%%%%%%%%%%%%%%%%%%%%%%%%%%
\end{document}